\documentclass[AMA,Times1COL]{WileyNJDv5}
\usepackage{graphicx}
\usepackage{subcaption}
\usepackage{amsmath, amssymb}
\usepackage{algorithm}
\usepackage{algorithmicx}
\usepackage{caption}
\usepackage{multirow}

\def\P{{\boldsymbol{P}}}
\def\C{{\boldsymbol{C}}}

\def\bSigma{{\boldsymbol{\Sigma}}}

\def\bgamma{{\boldsymbol{\gamma}}}

\articletype{Article Type}%

\received{Date Month Year}
\revised{Date Month Year}
\accepted{Date Month Year}
\journal{Journal}
\volume{00}
\copyyear{2023}
\startpage{1}

\begin{document}

\title{Multi-level Latent Variable Models for Coheritability Analysis in Electronic Health Records}

\author[1]{Yinjun Zhao }

\author[2,3]{Nicholas Tatonetti$^*$}

\author[1]{Yuanjia Wang$^*$}

\authormark{ZHAO \textsc{et al.}}
\titlemark{Multi-level Latent Variable Models for Coheritability Analysis in Electronic Health Records}

\address[1]{\orgdiv{Department of Biostatistics}, \orgname{Columbia University}, \orgaddress{\state{NY}, \country{USA}}}

\address[2]{\orgdiv{Department of Biomedical Informatics}, \orgname{Columbia University}, \orgaddress{\state{NY}, \country{USA}}}

\address[3]{\orgdiv{Department of Computational Biomedicine}, \orgname{Cedars-Sinai Medical Center}, \orgaddress{\state{CA}, \country{USA}}}

\corres{$^*$Co-corresponding authors: Yuanjia Wang, Nicholas Tatonetti \email{yw2016@cumc.columbia.edu, {nicholas.tatonetti@cshs.org}}}

\presentaddress{This is sample for present address text this is sample for present address text.}

%\fundingInfo{Text}
%\JELinfo{ejlje}

\abstract{Electronic health records (EHRs) linked with familial relationship data offer a unique opportunity to investigate the genetic architecture of complex phenotypes at scale. However, existing heritability and coheritability estimation methods often fail to account for the intricacies of familial correlation structures, heterogeneity across phenotype types, and computational scalability. We propose a robust and flexible statistical framework for jointly estimating heritability and genetic correlation among continuous and binary phenotypes in EHR-based family studies. Our approach builds on multi-level latent variable models to decompose phenotypic covariance into interpretable genetic and environmental components, incorporating both within- and between-family variations. We derive iteration algorithms based on generalized equation estimations (GEE) for estimation. Simulation studies under various parameter configurations demonstrate that our estimators are consistent and yield valid inference across a range of realistic settings. Applying our methods to real-world EHR data from a large, urban health system, we identify significant genetic correlations between mental health conditions and endocrine/metabolic phenotypes, supporting hypotheses of shared etiology. This work provides a scalable and rigorous framework for coheritability analysis in high-dimensional EHR data and facilitates the identification of shared genetic influences in complex disease networks.}

\jnlcitation{\cname{%
\author{Taylor M.},
\author{Lauritzen P},
\author{Erath C}, and
\author{Mittal R}}.
\ctitle{On simplifying ‘incremental remap’-based transport schemes.} \cjournal{\it J Comput Phys.} \cvol{2021;00(00):1--18}.}

\maketitle

\renewcommand\thefootnote{\fnsymbol{footnote}}
\setcounter{footnote}{1}

\label{sec1}
\section{Introduction}
%Significance of studying genetic contribution to comorbidity
Comorbidity, the co-occurrence of two or more disorders in a single individual, reflects the complex interactions that often exist between different disease processes. While lifestyle and environmental factors contribute to comorbidity, genetic factors, including pleiotropy, shared polygenic risk, and genetic correlation, play a foundational role in driving these patterns, particularly across complex diseases such as those in psychiatry \citep{co-heritability,smoller2016genetics}. Understanding the genetic basis of comorbidity is critical, as it may reveal shared etiological pathways, inform transdiagnostic classification frameworks, and ultimately guide the development of more targeted interventions.

%Despite growing recognition that many psychiatric and physical health conditions share common genetic underpinnings, most existing research has focused on individual disorders rather than the mechanisms underlying their co-occurrence \citep{valderas2009defining}.
Recent studies have highlighted substantial epidemiological comorbidity between mental and neurological disorders, such as depression, schizophrenia, and Parkinson’s disease, and metabolic, endocrine, and circulatory system diseases, including obesity, type 2 diabetes, and heart failure\citep{pan2012bidirectional, mitchell2013prevalence, gan2014depression, hu2007type, vancampfort2015risk, souza2021effect}. While some of these associations may be attributed to behavioral or lifestyle factors, a growing body of evidence points to an underlying shared genetic architecture across these disease domains.
Genome-wide association studies (GWAS) have revealed significant genetic correlations between several such phenotype pairs: major depressive disorder and body mass index (BMI) \citep{wray2018genome}, attention-deficit/hyperactivity disorder (ADHD) and type 2 diabetes \citep{baranova2023shared}, Alzheimer’s disease and type 2 diabetes \citep{hu2020shared}, and Parkinson’s disease and metabolic syndrome \citep{park2024multivariate}, among others.
To overcome the limitations of phenotype-specific approaches and advance the goals of precision medicine, this study aims to develop statistical methods for systematically estimating genetic correlations between mental and neurological diseases and metabolic, endocrine, and circulatory diseases. By analyzing the genetic underpinnings of comorbidity across these complex phenotypes, we aim to assist in understanding disease mechanisms and inform more effective strategies for risk prediction, prevention, and treatment.

%review of extisting literatures studying genetic correlation
Genetic overlap has traditionally been estimated using twin and family-based designs. Twin studies, which compare phenotypic concordance between monozygotic and dizygotic twins, provide a classical framework for partitioning variance into additive genetic and shared environmental components\citep{boomsma2002classical}. Family-based designs extend this framework by evaluating phenotypic resemblance among parents, offspring, and siblings to separate genetic influences from shared familial and individual-specific environmental effects \citep{jia2019estimating}. However, estimates from traditional twin studies or family-based studies may not be generalizable to the broader population due to ascertainment biases and limited diversity in family-based cohorts \citep{visscher2008heritability}.  Furthermore, it is challenging to scale these designs across a broad range of phenotypes, particularly for those that are rare, difficult to phenotype, or costly to measure \citep{bulik2015atlas}, which therefore cannot be used to study patterns of comorbidity at scale. 

\begin{figure}[ht]
  \centering
  \begin{subfigure}{0.8\textwidth}
    \includegraphics[width=\textwidth]{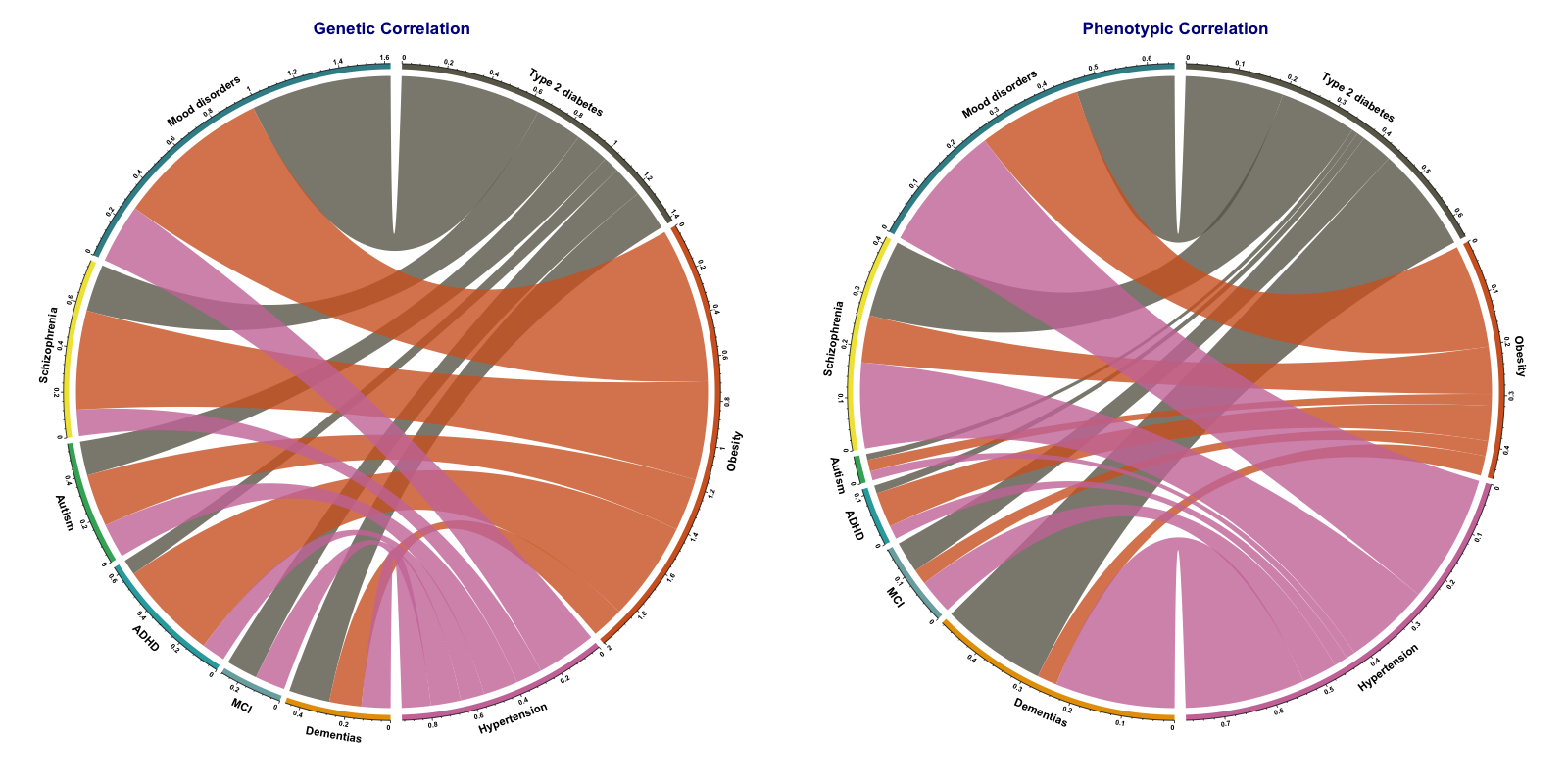}
    \caption{Binary phenotypes}
  \end{subfigure}
  \begin{subfigure}{0.8\textwidth}
    \includegraphics[width=\textwidth]{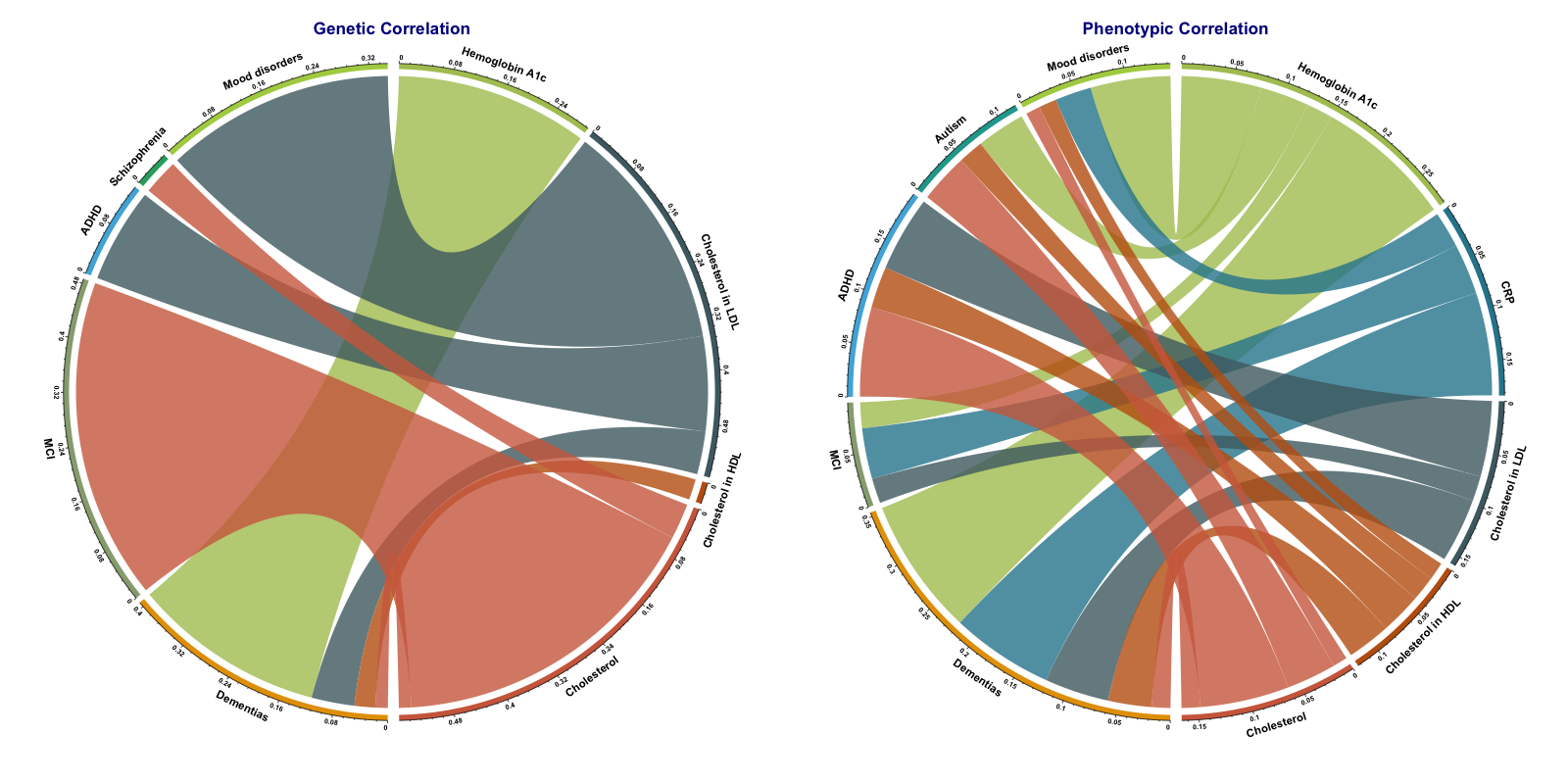}
    \caption{Mixed phenotypes}
  \end{subfigure}
  \caption{(a) Significant phenotypic and genetic correlations between binary mental/neurological disorders and binary endocrine/metabolic disorders, (b) Significant phenotypic and genetic correlations between one binary mental/neurological disorders and continuous endocrine/metabolic biomarkers; data is from
Columbia University Irving Medical Center in New York City EHRs; phenotypic correlations were calculated using polychoric correlation \citep{holgado2010polychoric} and genetic correlations were estimated by MPCE.}
  \label{fig:correlation}
\end{figure}

%More recently, the development of genome-wide association studies (GWAS) has enabled the estimation of heritability from common genetic variants in unrelated individuals. This approach, often referred to as SNP-based heritability, quantifies the proportion of phenotypic variance explained by the additive effects of common single-nucleotide polymorphisms (SNPs) measured in GWAS arrays. Methods such as LD Score regression \citep{bulik2015atlas} and genomic-relatedness-based restricted maximum likelihood (GREML)\citep{co-heritability} have been widely adopted to estimate SNP heritability using GWAS summary statistics or individual-level genotype data. In the SNP-based study, the relationship matrix captures information about just common SNPs rather than all genetic variation, so it provides a lower-bound estimate of total genetic contribution.
The growing availability of electronic health records (EHRs) has enabled researchers to access phenotypic data at an unprecedented scale, encompassing millions of individuals. 
%For example, the EHRs from Columbia University Irving Medical Center in New York City include more than 680,000 patients from 20 years of health records \cite{Polubriaginof}. 
However, because EHRs are primarily designed for administrative and billing purposes rather than research, they often suffer from limitations such as incompleteness and systematic biases\citep{jia2019estimating}. One notable limitation is the lack of explicit information on familial relationships, which restricts the ability to study genetic inheritance patterns. In addition, EHRs may not include all members of a nuclear family, further complicating family-based analyses.
To address this gap, Polubriaginof et al. developed the RIFTEHR (Relationship Inference From The Electronic Health Record) algorithm, which reconstructs familial relationships by linking emergency contact information in EHRs to other patients within the same healthcare system \citep{Polubriaginof}. By matching names, contact details, and declared relationship types, RIFTEHR inferred approximately 7.4 million first- to fourth-degree familial connections across three academic medical centers in New York City.
The availability of these reconstructed familial relationships presents an opportunity to assess phenotype correlations and genetic correlations using large-scale EHRs. For example, Figure~\ref{fig:correlation} shows significant polychoric correlations \citep{holgado2010polychoric} between neurological and psychiatric disorders and metabolic or endocrine conditions, using phenotypic data derived from the analyzed EHRs \citep{Polubriaginof}. These results reveal substantial covariation among several phenotype pairs, which may reflect shared genetic architecture, common environmental exposures, or both. Evidence of genetic correlation would support the hypothesis that these phenotypes share a common etiology.

To estimate the heritability of individual phenotypes and the co-inheritance of phenotype pairs, the ACE model is a widely used framework in family-based studies. It decomposes total phenotypic variance into three components: additive genetic variance (A), shared environmental variance (C), and non-shared environmental variance plus measurement error (E). By specifying a structured covariance matrix of phenotypes between family members, the ACE model captures the degree of genetic and environmental sharing between relatives \citep{Rabe-Hesketh}.
Parameters of the ACE model are typically estimated using one of several approaches: method-of-moments estimators such as Falconer’s equations \citep{falconer1996introduction}, structural equation modeling (SEM) \citep{bollen1989structural}, or likelihood-based methods assuming normally distributed phenotypes, commonly referred to as the normal ACE (NACE) model \citep{hesketh2008biometrical}. 

Although widely adopted in twin and family studies, the NACE model relies on the assumption of normally distributed phenotypes; violations of this assumption can lead to biased estimates and poor coverage of the true heritability parameters \citep{arbet2020robust}. SEM-based approaches similarly assume multivariate normality and can suffer decreased performance when this assumption is violated \citep{bollen1989structural}.
In contrast, moment-based estimators are generally more robust to deviations from normality, especially in large sample settings\citep{liang1986longitudinal}. To address these limitations, Arbet et al. \citep{arbet2020robust} proposed a robust generalized estimating equations framework (GEE2), which enables estimation of heritability for single phenotypes with non-normal distributions such as over-dispersed counts or heavy-tailed continuous outcomes. However, this approach is not applicable to jointly model heterogeneous data types.

For estimating heritability for binary phenotypes, a common model is the liability-threshold model, which assumes an underlying normally distributed latent liability. Individuals are considered to express the phenotype if their liability exceeds a certain threshold. Heritability is then estimated on the liability scale \citep{falconer1996introduction}. Likelihood-based estimation under this framework requires integration over the latent distribution and, due to the nonlinearity of the link function, does not offer explicit solutions as in the NACE model. Consequently, this approach incurs substantial computational costs, especially when integrating random effects in large-scale datasets such as those derived from EHRs \citep{mcculloch2004generalized}. Similarly, SEM is not well-suited for high-dimensional settings due to its memory and computational demands for non-normal phenotypes \citep{bollen1989structural}.
Due to these challenges, moment-based estimators offer a compelling alternative by avoiding the need for integration over individual-level random effects, thus providing greater computational efficiency and scalability for large, heterogeneous datasets.
Analogously to the framework by Arbet et al. \citep{arbet2020robust},  GEE can be applied to estimate the genetic correlation between two Gaussian-distributed phenotypes. However, no existing studies have jointly modeled phenotypes with heterogeneous data types, such as our focus on binary psychiatric diagnoses and continuous metabolic biomarkers. This gap highlights the need for developing methods to estimate genetic correlation and coheritability across mixed-type phenotypes in large-scale observational cohorts.    

%integration: need and challenge

In this paper, we develop a robust Multi-type Phenotype Co-inheritance Estimation method (MPCE) to estimate the genetic correlation of pairwise disorders with different data types. We propose multilevel linear mixed-effect models to jointly analyze phenotypes with heterogeneous data types and use a generalized method of moments estimator that can account for familial correlation and shared environmental factors but bypass high-dimensional integration or approximation. 
Our method tackles several challenges in estimating genetic correlation from EHRs. First, we address the challenge of a large sample size that results in a high-dimensional covariance matrix. Second, we account for different data types of phenotypes through latent liability variables. Third, we account for the multi-level dependence structure between comorbid disorders within a subject and the multi-level genetic and environmental correlation structure between subjects within a family. Lastly, our method accounts for missing phenotypes and incomplete family members by inverse probability weighting based on informative covariates available in EHRs.

We conducted extensive simulations to examine the consistency and inference of the proposed method. Comparisons with univariate methods were used to evaluate estimation accuracy, efficiency gain, and robustness under different scenarios, and the sensitivity of our method to various assumptions was also assessed. We applied the proposed methods to analyze EHRs from Columbia University Irving Medical Center in New York City. Using our method, we estimated genetic correlations between neurological and psychiatric disorders, as well as various metabolic disorders and biomarkers. Our results validate existing literature and provide new perspectives for studying the mechanism of their associations and, therefore, facilitate developing effective and tailored treatments.

%Moreover, we propose a computationally and memory-efficient multilevel method to improve power for detecting genetic variants associated with the risk of multivariate phenotypes in the combined data of proband genotypes and family history in relatives (PheWAS+FH).

\section{Methods}
\label{sec2}

\subsection{Model}
Let $Y_{ijk}$ denote the $k$th phenotype in the $j$th subject from the $i$th family for $i=1,\cdots,n$, $j=1,\cdots, n_i$, $k=1,\cdots,K$. For multivariate continuous phenotypes, a multilevel linear mixed effects model is 
\begin{equation}
\label{eq1}
    Y_{ijk}=\alpha_k + \boldsymbol{\beta_k}^T \boldsymbol{X_{ijk}} + \gamma_kb_{i} + u_{ijk}  +\epsilon_{ijk},
\end{equation}
where $b_i$ is the shared random effect due to shared environment (e.g., lifestyle, diet), $\gamma_k$ reflects how this shared environment relates to disease $k$ with $\gamma_1=1$ for identifiability, $u_{ijk}$ denotes the random effects due to family-inherited disease-specific genes, and $\epsilon_{ijk}$ is measurement error assumed to follow $N(0,\sigma_{\epsilon k}^2)$.
For $u_{ijk}$, we assume that for any $j$, 
$$\mbox{cov} \left(\begin{array}{c} u_{ij1}\\ \vdots \\ u_{ijK}\end{array}\right)=\left(
\begin{array}{cccc}
\sigma^2_1 & \rho_{12}\sigma_1\sigma_2 & \cdots & \rho_{1K}\sigma_1\sigma_K\\
\rho_{12}\sigma_1\sigma_2 & \sigma^2_2 & \cdots & \rho_{2K}\sigma_2\sigma_K \\
\vdots & \vdots &\vdots &\vdots \\
\rho_{1K}\sigma_1\sigma_K & \rho_{2K}\sigma_2\sigma_K & \cdots &\sigma_K^2
\end{array}
\right)_{K\times K},$$
denoted as $\bSigma$, and that for $j\neq s$, $\mbox{cov}(u_{ijk}, u_{isk})=\sigma_k^2 c_{ijs}$ and $\mbox{cov}(u_{ijk}, u_{ism})=\rho_{km}\sigma_k\sigma_{m} c_{ijs}$,
where 
$c_{ijs}$ is the kinship coefficient between individuals $j$ and $s$ in family $i$. If SNP genotypes are collected from unrelated subjects, $c_{ijs}$ can be estimated from identical-by-descent (IBD) sharing or a genetic relationship matrix \citep{yang2011gcta} using the genotypes. 
Equivalently, the random vector of $\boldsymbol{U}=(u_{i1k},\cdots,u_{in_i k}, k=1,\cdots,K)$ has the covariance matrix
$$\mbox{cov}(\boldsymbol{U})=\bSigma\otimes \C_i, \ \ \C_i=(c_{ijs})_{n_i\times n_i}.$$ Note that the genetic inheritance of a single phenotype between subjects ($j,s$) within the same family is determined by $\C_i=(c_{ijs})$ and diagonal terms of $\bSigma$: $\mbox{cov}(u_{ijk},u_{isk}) = c_{ijs}\sigma_k^2$. In addition, the coinheritance of different phenotypes for related subjects within a pedigree identifies the genetic coinheritance through off-diagonal terms of $\bSigma$: $\mbox{cov}(u_{ijk},u_{isk'}) = c_{ijs}\sigma_k\sigma_{k'}$.
Thus, for the observed vector of phenotypes $\boldsymbol{Y}_i=(Y_{i1k},\cdots,Y_{in_i k}, k=1,\cdots,K)$, 
 \begin{eqnarray*}
      \mbox{cov} (\boldsymbol{Y}_i)=\sigma_b^2 (\bgamma\bgamma^T)\otimes \boldsymbol{J}\boldsymbol{J}^T+\bSigma\otimes \C_i+\textrm{diag}(\sigma_{\epsilon 1}^2,...,\sigma_{\epsilon K}^2)\otimes \boldsymbol{I}_i,
 \end{eqnarray*}
where $\bgamma=(\gamma_1,...,\gamma_K)^T$, $\boldsymbol{J}$ denotes $(1,...,1)_{n_i}^T$ and $\boldsymbol{I}_i$ is $n_i\times n_i$ identity matrix.
For identifiability, we assume $\gamma_1=1$.
Specifically, when there is a single phenotype and $K=1$, 
 \begin{eqnarray*}
  \mbox{cov} (\boldsymbol{Y}_i)=\sigma_b^2  \boldsymbol{J}\boldsymbol{J}^T+\sigma_1^2 \C_i+\sigma_{\epsilon 1}^2 \boldsymbol{I}_i,   
 \end{eqnarray*}
and our model reduces to the usual model used to estimate heritability.

To generalize to dichotomous phenotypes where $Z_{ijk}$ are binary $\{0,1\}$, we follow the underlying variable (UV) approach commonly used in social science \citep{bartholomew2008analysis} to assume that there are 
continuous latent liability variables $Y_{ijk}$ such that $Z_{ijk}=I(Y_{ijk}>0)$ and $Y_{ijk}$'s satisfy the same models as in \eqref{eq1}. 
Under this framework, we can integrate continuous and binary phenotypes, where the genetic heritability is defined through latent liabilities. With the variance components estimated, the heritability of the individual univariate phenotype for phenotype $k$, $h^2_k$, is calculated by the proportion of genetic variation for phenotype $k$ among the total variation $\displaystyle\frac{\sigma^2_k}{\sigma^2_k+\sigma^2_b+\sigma^2_{\epsilon_k}}$.
The coheritability between phenotype $k$ and $m$, denoted by $h^2_{km}$, is computed as $$h_{km}^2=\rho_{km} h_k h_m,$$ where $\rho_{km}$ is the genetic correlation coefficient between phenotype $k$ and phenotype $m$, $h_k$ and $h_m$ are the root heritability \citep{falconer1996introduction}.

\subsection{Estimation and Algorithm}
Model \eqref{eq1} is usually fit by restricted maximum likelihood. However, when the sample size is large, such as EHR data (e.g., over 40K families with 10K subjects), the covariance matrix for $\boldsymbol{Y}$ is high-dimensional with high-dimensional random effects. Thus, algorithms for the mixed effects model (especially for binary phenotypes) would face computational challenges in integrating over high-dimensional random effects. We propose an alternative method based on generalized methods of moments (GMM) \citep{liang1986longitudinal}, which is computationally efficient and offers consistent estimators. We first use GMM to estimate $\boldsymbol\beta$ and $\boldsymbol\theta$ for the multivariate continuous phenotypes model and then extend to binary phenotypes and phenotypes of different data types. Denote $\boldsymbol\beta=\left(\alpha_1, \boldsymbol{\beta_1}^T, \dots, \alpha_K, \boldsymbol{\beta_K}^T\right)$, and $\boldsymbol{\theta}=\left(\gamma_2,\dots\gamma_k, \sigma_1\dots \sigma_K, \sigma_b,  \sigma_{\epsilon_1}\dots\sigma_{\epsilon_K}, \{\rho_{km}\}_{k,m=1,\cdots,K,k\neq m}\right)$.  The iterative algorithms for the multivariate continuous phenotypes model are presented in Algorithm \ref{alg:continuous}. 
The first step is to initialize the sample covariance of outcomes $\widehat{\text{cov}}(\boldsymbol{Y})$ using residuals from the mean.
The second step is to iteratively update $\boldsymbol{\hat\beta}$  using a GEE solution for $\boldsymbol{\hat\beta}$ weighted by the inverse of the current estimate of the covariance matrix,
and re-estimate the residual covariance matrix and $\widehat{\text{cov}}(\boldsymbol{Y})$ using the updated $\boldsymbol{\hat\beta}$ until $\boldsymbol{\hat\beta}$ converges.
The third step is to estimate covariance parameters $\boldsymbol{\hat\theta}$ by solving the equation system $V(\boldsymbol{\theta}) = \widehat{\text{cov}}(\boldsymbol{Y})$,
and to estimate $\boldsymbol{\hat\beta}$ one final time using $V(\boldsymbol{\hat\theta})$ as the known covariance structure.

\begin{algorithm}
\caption{Algorithm for multivariate continuous phenotypes model}\label{alg:continuous}
\begin{algorithmic}[htb]
\State \textbf{Data}: covariates $\boldsymbol X$, outcomes $\boldsymbol Y$
\State \textbf{Results}: Estimators $\boldsymbol{\hat\beta}$, $\boldsymbol{\hat\theta}$
\State \textbf{Step 1}: Initialize $\widehat{\text{cov}}(\boldsymbol Y) = \frac{1}{n} \sum_{i=1}^n \left( \boldsymbol Y_i - \bar{\boldsymbol Y_i} \right) \left( \boldsymbol Y_i - \bar{\boldsymbol Y_i} \right)^T$
\State \textbf{Step 2}: Repeat
\State \quad  $\boldsymbol{\hat\beta} = \left[ \sum_{i=1}^n \boldsymbol X_i^T [ \widehat{\text{cov}}(\boldsymbol Y) ]^{-1} \boldsymbol X_i \right]^{-1} \left[ \sum_{i=1}^n \boldsymbol X_i^T [ \widehat{\text{cov}}(\boldsymbol Y) ]^{-1} \boldsymbol Y_i \right]$
 \State   \quad $\widehat{\text{cov}}(\boldsymbol Y) = \frac{1}{n} \sum_{i=1}^n \left( \boldsymbol Y_i - \boldsymbol X_i \boldsymbol{\hat\beta} \right) \left( \boldsymbol Y_i - \boldsymbol X_i \boldsymbol{\hat\beta} \right)^T$
\State \quad Until $\boldsymbol{\hat\beta}$ converges
\State \textbf{Step 3}: update $\boldsymbol{\hat\theta}$ by solving $V(\boldsymbol\theta) = \widehat{\text{cov}}(\boldsymbol Y)$
\State \textbf{Step 4}: update $\boldsymbol{\hat\beta} = \left[ \sum_{i=1}^n \boldsymbol X_i^T [ V(\boldsymbol{\hat\theta}) ]^{-1} \boldsymbol X_i \right]^{-1} \left[ \sum_{i=1}^n \boldsymbol X_i^T [ V(\boldsymbol{\hat\theta}) ]^{-1} \boldsymbol Y_i \right]$
\end{algorithmic}
\end{algorithm}

We provide an example of two continuous phenotypes for nuclear families with two parents and two children, i.e., $K=2, n_i=4$, for each $i=1, \ldots, n$, to illustrate how to solve the equation system. Suppose that the kinship matrix is 
$$C=\left( \begin{matrix}
    1 &0 &1/2 &1/2 \\
    0 & 1 & 1/2 & 1/2\\
    1/2 & 1/2 & 1 & 1/2 \\
    1/2 & 1/2 & 1/2 &1\\
\end{matrix}\right). $$
Then components of $cov(\mathbf Y)$ can be expressed as
\begin{eqnarray*}
var(Y_{\cdot j1})&=&\gamma_1^2\sigma_b^2+\sigma_1^2+\sigma_{\epsilon 1}^2, \quad 
var(Y_{\cdot j2})=\gamma_2^2\sigma_b^2+\sigma_2^2+\sigma_{\epsilon 2}^2, \quad cov(Y_{\cdot j1}, Y_{\cdot j2})=\gamma_1\gamma_2\sigma_b^2+\rho_{12}\sigma_1\sigma_2,
\end{eqnarray*}
\begin{eqnarray*}
%cov(Y_{\cdot j1}, Y_{\cdot j2})&=&\gamma_1\gamma_2\sigma_b^2+\rho_{12}\sigma_1\sigma_2\\
cov(Y_{\cdot j1}, Y_{\cdot s1})&=&
\begin{cases}
\gamma_1^2\sigma_b^2+1/2 \sigma_1^2, (j,s)\in \mathcal{A}_1 \\
\gamma_1^2\sigma_b^2,(j,s) \in \mathcal{A}_0
\end{cases}\\
cov(Y_{\cdot j2}, Y_{\cdot s2})&=&
\begin{cases}
\gamma_2^2\sigma_b^2+ 1/2 \sigma_2^2, (j,s)\in \mathcal{A}_1 \\
\gamma_2^2\sigma_b^2,(j,s)\in \mathcal{A}_0
\end{cases}\\
cov(Y_{\cdot j1}, Y_{\cdot s2})&=&
\begin{cases}
\gamma_1\gamma_2\sigma_b^2+ 1/2\rho_{12}\sigma_1\sigma_2, (j,s)\in \mathcal{A}_1\\
\gamma_1\gamma_2\sigma_b^2, (j,s)\in \mathcal{A}_0
\end{cases},
\end{eqnarray*}
where $\mathcal{A}_1=\{(j,s):c_{js}=1/2\}, 
\mathcal{A}_0=\{(j,s):c_{js}=0\}, j\neq s, j=1\cdots,4$. 
The equation system has 9 equations with 6 parameters $(\gamma_2, \sigma_1, \sigma_2, \sigma_b,  \sigma_{\epsilon_1}\sigma_{\epsilon_2})$.

Let $$
a_1 =\frac{1}{4} \sum_{j=1} ^{4} \widehat{Var(Y_{\cdot j1})}, a_2 =\frac{1}{4} \sum_{j=1} ^{4} \widehat{Var(Y_{\cdot j2})},a_3 =\frac{1}{4} \sum_{j=1} ^{4} \widehat{Cov(Y_{\cdot j1}, Y_{\cdot j2})},$$
$$a_4 =\frac{1}{5}\sum_{(j,s)\in \mathcal{A}_1} \widehat{Cov(Y_{\cdot j1}, Y_{\cdot s1})},
a_5 = \widehat{Cov(Y_{\cdot j1}, Y_{\cdot s1})} ,{(j,s)\in \mathcal{A}_0},$$
$$a_6 =\frac{1}{5}\sum_{(j,s)\in \mathcal{A}_1} \widehat{Cov(Y_{\cdot j2}, Y_{\cdot s2})},
a_7 = \widehat{Cov(Y_{\cdot j2}, Y_{\cdot s2})},{(j,s)\in \mathcal{A}_0},$$
$$a_{8} =\frac{1}{5}\sum_{(j,s)\in \mathcal{A}_1} \widehat{Cov(Y_{\cdot j1}, Y_{\cdot s2})},
a_{9} = \widehat{Cov(Y_{\cdot j1}, Y_{\cdot s2})}, {(j,s)\in \mathcal{A}_0}.$$
Solving this equation system can be converted into minimizing the objective function 
\begin{align*}
f(\sigma_1, \sigma_2, \sigma_{b_1}, \sigma_{b_2}, \sigma_{\epsilon_1}, \sigma_{\epsilon_2}) 
= \; & 4(\sigma_{b_1}^2+\sigma_1^2+\sigma_{\epsilon 1}^2-a_1)^2
+ 4(\sigma_{b_2}^2+\sigma_2^2+\sigma_{\epsilon 2}^2-a_2)^2 \\
& + 4(\sigma_{b_1}\sigma_{b_2}+\rho_{12}\sigma_1\sigma_2-a_3)^2
+ 5(\sigma_{b_1}^2+\tfrac{1}{2} \sigma_1^2-a_4)^2 
+ (\sigma_{b_1}^2-a_5)^2 \\
& + 5(\sigma_{b_2}^2+\tfrac{1}{2} \sigma_2^2-a_6)^2
+ (\sigma_{b_2}^2-a_7)^2
+ 5(\sigma_{b_1}\sigma_{b_2}+ \tfrac{1}{2}\rho_{12}\sigma_1\sigma_2-a_8)^2 \\
& + (\sigma_{b_1}\sigma_{b_2} -a_9)^2.
\end{align*}
Next, for the multivariate binary phenotypes model, we use the same method of moments to construct generalized estimating equations for all parameters. The estimating equations involve the first or second interactions among $Y_{ijk}$'s. Since $Y_{ijk}$ are latent variables, they are replaced with conditional expectations given observed binary phenotypes,
$E\left[Y_{ijk}|Z_{ijk}, \boldsymbol {X_i}\right]$ and $E\left[Y_{ijk}Y_{ism}|Z_{ijk}, Z_{ism}\right]$, respectively, which can be computed numerically based on integration of at most two-dimensional multivariate normal distribution. This yields the iterative algorithms shown in Algorithm \ref{alg:binary}. 
The first step is to initialize $\boldsymbol{\hat\beta}$ using a probit regression assuming subjects are independent and
	 $\boldsymbol{\hat\theta}$ using GMM, treating binary outcomes as continuous.
The second step is to iteratively update the model until convergence: a) Update the latent covariance matrix $\widehat{\text{cov}}(\boldsymbol{Y})$, where $\boldsymbol{Y}_i$ represents the latent variables underlying $\boldsymbol{Z}_i$. b) Solve the equation system $V(\boldsymbol{\theta}) = \widehat{\text{cov}}(\boldsymbol{Y})$ to update $\boldsymbol{\hat\theta}$.
c) Update $\boldsymbol{\hat\beta}$ using GEE estimators, where the outcome is replaced by the conditional expectation $\mathbb{E}[\boldsymbol{Y}_i \mid \boldsymbol{X}_i, \boldsymbol{Z}_i, \boldsymbol{\hat\theta}, \boldsymbol{\hat\beta}]$.

\begin{algorithm}[hbt!]
\caption{Algorithm for multivariate binary phenotypes model}\label{alg:binary}
\begin{algorithmic}
\State \textbf{Data}: Covariates $\boldsymbol X$, outcomes $\boldsymbol Z$
\State \textbf{Results}: Estimators $\boldsymbol{\hat\beta}$, $\boldsymbol{\hat\theta}$
\State \textbf{Step 1} Initialize 
\State \quad$\boldsymbol{\hat\beta}$ using probit model assuming all the subjects are i.i.ds
\State \quad $\boldsymbol{\hat\theta}$ using GMM treating observed $\boldsymbol Z$ as $\boldsymbol Y$ 
 \State \textbf{Step 2}: Repeat
\State \quad $\widehat{cov(\boldsymbol Y)}$
\State \quad
     $\boldsymbol{\hat\theta}$ by solving the equation system $V(\boldsymbol\theta)=\widehat{cov(\boldsymbol Y)}$
\State \quad $\boldsymbol{\hat\beta}=\left[\sum_{i=1}^n \boldsymbol X_i^{'}[V(\boldsymbol {\hat\theta})]^{-1}\boldsymbol X_i\right]^{-1} \left[\sum_{i=1}^n \boldsymbol X_i^{'}[V(\boldsymbol{\hat\theta})]^{-1}\boldsymbol Y_i\right]$, where \textcolor{black}{$\boldsymbol Y_i$ is replaced by $E(\boldsymbol Y_i|\boldsymbol X_i,\boldsymbol Z_i, \boldsymbol{\hat\theta}, \boldsymbol{\hat\beta})$}
  \State \quad Until $\boldsymbol{\hat\beta}$ and $\boldsymbol{\hat\theta}$ converge 
 \end{algorithmic}
\end{algorithm}

To estimate $cov(\boldsymbol{Y})$, we expand $cov(Y_{\cdot jk}, Y_{\cdot sm})$ into 
$$E\left(cov(Y_{\cdot jk}, Y_{\cdot sm}|X_{\cdot jk}, X_{\cdot sm}, Z_{\cdot jk}, Z_{\cdot sm})\right)+
cov\left(E(Y_{\cdot jk}|X_{\cdot jk}, Z_{\cdot jk},Z_{\cdot sm}), E(Y_{\cdot sm}|X_{\cdot sm}, Z_{\cdot jk}, Z_{\cdot sm})\right),$$ 
which equals the sum of two expectations, that is, $$
E\left[
    E\left( Y_{\cdot jk}, Y_{\cdot sm} \mid X_{\cdot jk}, X_{\cdot sm}, Z_{\cdot jk}, Z_{\cdot sm} \right)
    - E\left( Y_{\cdot jk} \mid X_{\cdot jk}, Z_{\cdot jk}, Z_{\cdot sm} \right) \cdot X_{\cdot sm} \beta_{sm}
    - E\left( Y_{\cdot sm} \mid X_{\cdot sm}, Z_{\cdot jk}, Z_{\cdot sm} \right) \cdot X_{\cdot jk} \beta_{jk}
\right]
$$
and
$E\left[X_{\cdot jk} \beta_{jk} X_{\cdot sm} \beta_{sm}\right].$
%We denote these two expectations as $(I)$ and $(II)$, respectively. 
We estimate the first term by the sample mean of 
$E(Y_{ijk}, Y_{i sm}|X_{ijk}, X_{ism}, Z_{ik}, Z_{ism})-E(Y_{ijk}|X_{ijk}, Z_{i jk}, Z_{ism})*X_{ism} \beta_{sm}
-E(Y_{ism}|X_{ism}, Z_{i jk}, Z_{i sm})*X_{ijk} \beta_{jk},$
and estimate the second term by the sample mean of $X_{ijk} \beta_{jk} X_{ism} \beta_{sm}, i=1, \cdots, n$. 
By handling a high-dimensional covariance matrix in a pairwise fashion and using GMM, we avoid the numerical difficulty of high-dimensional integration of random effects.

To improve the efficiency of iterations, we first fix $\boldsymbol{\hat\beta} $ and update $\boldsymbol{\hat\theta}$ until it converges, and then update $\boldsymbol{\hat\beta}$ and $\boldsymbol{\hat\theta}$ simultaneously until they both converge. This algorithm applies when some family members are missing or not all the families have the same family size. We achieve this by keeping their family's covariates matrix's dimension the same as the complete family's and treat missings as zero since they don't contribute to estimating $\boldsymbol \beta$.

The algorithms also apply when the phenotypes include both continuous and binary phenotypes. Suppose that the $k$th outcome is continuous, and the $m$th outcome is binary, i.e., $Y_{\cdot jk}, Z_{\cdot sm}$ is observed, and $Y_{\cdot sm}$ is latent. We expand $cov(Y_{\cdot jk}, Y_{\cdot sm})$ into
$$E\left(cov(Y_{\cdot jk}, Y_{\cdot sm}|X_{\cdot jk}, X_{\cdot sm}, Z_{\cdot sm})\right)+cov\left(E(Y_{\cdot jk}|X_{\cdot jk}, Z_{\cdot sm}), E(Y_{\cdot sm}|X_{\cdot sm}, Z_{\cdot sm})\right),$$ which equals sum of two expectations, i.e., 
$$E\left[E(Y_{\cdot jk}, Y_{\cdot sm}|X_{\cdot jk}, X_{\cdot sm}, Z_{\cdot sm})-
E(Y_{\cdot jk}|X_{\cdot jk}, Z_{\cdot sm})*X_{\cdot sm} \beta_{sm}
-E(Y_{\cdot sm}|X_{\cdot sm}, Z_{\cdot sm})*X_{\cdot jk} \beta_{jk}\right]$$
and $E\left[X_{\cdot jk} \beta_{jk} X_{\cdot sm} \beta_{sm}\right].$ The first expectation can be estimated by the sample mean of $E(Y_{ijk}, Y_{i sm}|X_{ijk}, X_{ism}, Z_{ism})-
E(Y_{ijk}|X_{ijk}, Z_{i sm})*X_{ism} \beta_{sm}
-E(Y_{ism}|X_{ism}, Z_{i sm})*X_{ijk} \beta_{jk}$, and the second expectation is the same as for the binary phenotypes. 

This algorithm also works when the families are from stratified sampling or when families have unequal weights. In this case, $\boldsymbol{\hat\theta}$ is initialized by weighted GMM, $\frac{1}{n}\sum _{i=1}^n w_i \left(\boldsymbol Y_{i}-\bar {\boldsymbol Y}\right)\left(\boldsymbol Y_{i}-\bar{\boldsymbol Y}\right)^\top$, using pairwisely complete observations, where $\bar{\boldsymbol Y}$ is the weighted average,  $\boldsymbol{\hat\beta}$ is updated by $$\left[\sum_{i=1}^n w_i \boldsymbol X_i^{'}\left[V\left(\boldsymbol{\hat\theta}\right)\right]^{-1}\boldsymbol X_i\right]^{-1} \left[\sum_{i=1}^n w_i \boldsymbol X_i^{'}\left[V\left(\boldsymbol{\hat\theta}\right)\right]^{-1}\boldsymbol Y_i\right],$$ and $\mbox{cov}(\boldsymbol Y)$ is estimated by the weighted sample mean.

%\subsection{Inference}
We use parametric bootstrap to obtain the inference of all the parameters. For the multivariate continuous phenotypes model, we first apply Algorithm \ref{alg:continuous} to the data $\left(\boldsymbol X, \boldsymbol Y\right)$ to obtain the estimation of all parameters $\left(\boldsymbol {\hat \beta}^*, \boldsymbol  {\hat\theta}^*\right)$. Second, we randomly sample from $N\left( \boldsymbol X\boldsymbol {\hat\beta}^*, V\left(\boldsymbol {\hat \theta}^*\right)\right)$ to obtain $\boldsymbol{Y^*}$ and further generate bootstrap samples $(\boldsymbol X, \boldsymbol Y^*)$. Third, we apply the same algorithm to the bootstrap samples and obtain estimated values $\left(\boldsymbol {\hat\beta}, \boldsymbol {\hat\theta}\right)$. Last, we repeat the second and third steps $N$ times to obtain 95\% CI of $\left(\boldsymbol {\hat\beta}, \boldsymbol {\hat\theta}\right)$ and their variances. 

For the multivariate binary phenotypes model, we apply Algorithm \ref{alg:binary} to the data $(\boldsymbol X, \boldsymbol Z)$ to obtain $\left(\boldsymbol {\hat\beta}^*, \boldsymbol {\hat\theta}^*\right)$ and then apply the same algorithm to the bootstrap samples $\left(\boldsymbol X, \boldsymbol Z^*\right)$ to obtain the estimated values $\left(\boldsymbol {\hat\beta}, \boldsymbol {\hat\theta}\right)$. In generating the bootstrap samples $(\boldsymbol X, \boldsymbol Z^*)$, we randomly sample $\boldsymbol Y^*$ from $N\left(\boldsymbol X\boldsymbol {\hat\beta}^*, V\left(\boldsymbol {\hat\theta}^*\right)\right)$ and dichotomize $\boldsymbol Y^*$ to be $\boldsymbol Z^*$ by 
$Z^*_{ijk}=I\left(Y^*_{ijk}>0\right)$. For the multivariate mixed phenotypes model, a similar approach applies. When generating bootstrap samples, discretization is applied when the phenotype is binary.

\section{Simulation studies}
\label{sec3}
We conducted comprehensive simulation studies to evaluate the performance of the methods for the multivariate continuous phenotypes model, the multivariate binary phenotypes model, and the multivariate mixed phenotypes model, respectively. We examined estimation consistency, inference, and coverage, and compared them to simpler approaches. Furthermore, we conducted a sensitivity analysis to examine the robustness of our methods. 

%\subsection{Consistency of the estimators}

To evaluate the performance of the multivariate continuous phenotypes model, we conducted simulations with two continuous phenotypes ($K =2$). We refer to this scenario as a two-continuous-phenotype analysis. Similarly, we use two-binary-phenotype and two-mixed-phenotype analyses to denote the simulation scenarios with two binary outcomes, one continuous outcome, and one binary outcome, respectively. We simulated four members in each family (i.e., $n_i=4$ for all $i$), 
and the kinship matrix is 
$$C=\left( \begin{matrix}
    1 &0 &1/2 &1/2 \\
    0 & 1 & 1/2 & 1/2\\
    1/2 & 1/2 & 1 & 1/2 \\
    1/2 & 1/2 & 1/2 &1\\
\end{matrix}\right). $$ 

\vspace{5mm}

Four covariates were included $X_1, X_2, X_3, X_4$ in the model, where $X_1$ and $X_2$ followed binomial distributions, and $X_3$ and $X_4$ followed a normal distribution and an exponential distribution, respectively. We randomly sampled true values of the parameters $\boldsymbol{\beta}, \boldsymbol{\theta}$ from standard uniform distributions. 
To accommodate the varying levels of heritability and genetic correlation across different phenotypes, we considered four groups of settings with different combinations of the heritability of phenotype 1, phenotype 2,  and the genetic correlation between these two phenotypes as: $(h_1^2, h_2^2, \rho)=(0.61, 0.54, 0.3), (0.54, 0.35, 0.3), (0.46, 0.35, 0.3), (0.61, 0.54, 0.6)$. We denote these settings as (high, high, low), (high, low, low), (low, low, low), (high, high, high), respectively. 

We sampled the continuous outcome $\boldsymbol{Y}$ from $N\left(\boldsymbol X \boldsymbol \beta, V\left(\boldsymbol \theta\right)\right)$.  We conducted 100 replications for each group setting with 500 or 1000 families.
Figure \ref{fig:continuous,high,high,low} shows the simulation results for the (high, high, low) setting, where the heritability of phenotype 1 and phenotype 2 are both high, and the genetic correlation is low. The figure shows that the medians of the estimated values are close to the true values, and as the number of families increases to 1000, the variance of the estimators decreases. We compared the root mean squared errors (RMSE) of all parameters for 500 families and 
1000 families. As shown in 
Table \ref{table:RMSE}, all parameters' RMSEs decrease as the family size increases from 500 to 1000 with an average decrease of 28\% for the (high, high, low) group. The other three settings have similar performance, and results are omitted. 

In simulation studies for two-binary-phenotype analysis, the same covariates $\boldsymbol{X}$ and the parameters $\boldsymbol{\beta}$, $\boldsymbol {\theta}$ were used as the two-continuous-phenotype analysis. For the two-binary-phenotype analysis, the phenotype values $\boldsymbol{Y}$ were dichotomized by applying a threshold of zero, i.e.,  $Z_{ijk}=I\left(Y_{ijk}>0\right), k=1,2$. Similarly, we used the same covariates and true parameters for the two-mixed-phenotype analysis. The values of its first phenotype were dichotomized as  $Z_{ij1}=I\left(Y_{ij1}>0\right)$, and the second phenotype was the same as the second phenotype in the two-continuous-phenotype analysis (i.e., $Y_{ij2}$). As shown in Figures \ref{fig:binary,high,high,low} and \ref{fig:mixed,high,high,low}, the medians of the estimated values in both analyses are close to their corresponding true values, and the variation decreases as the number of families increases. 
In the two-binary-phenotype analysis and the two-mixed-phenotype analysis for the (high, high, low) setting, 
Table \ref{table:RMSE}) shows that all parameters’ RMSEs decrease as the number of families increases from 500 to 1000, with an average decrease
of 34.2\% and 24\%, respectively. The other three settings have similar results. %The results show that the estimated values are consistent.
%The box plots of the simulation results for heritability high \&low, low\&low settings for the above three types of two-phenotype analysis are in the appendix. 

%\subsection{Inference using parametric bootstrap}

Parametric bootstrap was used to obtain the 95\% confidence intervals (CIs) for all the parameters in $\boldsymbol \beta$ and $\boldsymbol \theta$. We repeated this procedure $M$ times to obtain the 95\% CI coverage rate for each parameter. 
In the simulation studies, we set $M$ to 100 and used the same settings of parameters and covariates in the two-phenotype analysis. As shown in Table \ref{table:ci rate}, the 95\% CI coverage rates for all parameters are around 0.95, except for $\rho$, which has a coverage rate of one. This conservative coverage rate may be due to numerical issues in solving the covariance equation system.


%\subsection{Comparison with a simpler approach: single-phenotype analysis}

We conducted a single-phenotype analysis for each of the two phenotypes separately, using the same settings of  parameters and covariates as in the two-phenotype analysis. We compared the performance of  the single-phenotype analysis with the two-phenotype analysis ('integrated')  by examining the RMSE percent difference. As shown in Table \ref{table:RMSE}, the single-phenotype analysis ('separated'), tends to have higher RMSEs than the two-phenotype analysis ('integrated') for all three types of two-phenotype integration except for $\boldsymbol{\beta}'s$ in the two-binary-phenotype integration and the two-mixed-phenotype integration. Across the three heritability settings, the RMSE improves 6.9\%, 6.9\% and 0.9\% on average in the three types of integration (two-continuous, two-binary, and two-mixed), respectively. These results show that integrating multivariate phenotypes leads to more efficient parameter estimation than analyzing each phenotype separately. 

%\subsection{Sensitivity Analysis}
A sensitivity analysis was conducted to consider reporting errors where: 1) families are not independent, such as when one parent from family A and one parent from family B are siblings; or 2) children from the same families are not biologically related. To measure the impact of reporting errors on the accuracy of the two-phenotypes analysis, we set the reporting error rates to be 5\% and 10\%, respectively, and calculated the RMSE percent difference from families with and without reporting errors. As shown in Table \ref{table:RMSE}, the RMSE average percentage difference ranges from $-6.8\%$ to $8.7\%$ for the 5\% reporting error rate, and from $-2.3\%$ to $7.4\%$ for the 10\% reporting error rate. This sensitivity analysis shows that our approach is robust to a reporting error of 5\% or 10\%.

\section{Analysis of the EHR data}
We applied MPCE to the EHR data with next-of-kin information collected through patient emergency contact data from Columbia University Irving Medical Center in New York City. These data include 680,000 patients and 3.2 million relationships for 223,000 families inferred from 20 years of health records \cite{Polubriaginof}.  The familial relationships were initially established by the patients whose emergency contact had a family relationship with the patient and were also patients in the EHRs, resulting in 488,932 relationships. Then, through the next-of-kin relationship, an additional 2,755,448 relationships were inferred. \cite{Polubriaginof} evaluated the inferred relationships using documented mother-child relationships for newborns delivered in hospitals. They showed 92.9\% sensitivity and 95.7\% positive predictive value (PPV) for maternal
 relationships and 92.2\% sensitivity and 98.3\% PPV for siblings.
 
Diseased subjects were identified as individuals with a certain disorder (e.g., Parkinson's disease, PD) from the database using the Phecode map \citep{phecode} and non-diseased subjects were identified by excluding individuals with the disorder of interest and excluding individuals who had closely related disorders (e.g., other cerebral degeneration, disorders of the autonomic nervous system).
The data processing steps were illustrated using PD as an example, as in Fig \ref{fig:continuous,high,high,low}. First, patients who had observed values for Parkinson’s disease ICD codes or any other ICD codes out of all hospitalized patients were identified.
PD patients were defined as patients with Phecode 332 that mapped to ICD-9 code 332 or ICD-10 code G21.4 or G20 and non-PD patients were defined as individuals who were neither PD nor had  Phecode in the range of conditions closely related to Parkinson's disease (e.g., 330-337.99, 341-349.99). Second, individuals' family members were identified using the familial relationships provided in Polubriaginof et al.\citep{Polubriaginof}. 

We focused on nuclear families, which comprised the majority of relationships (1,388,858 out of 3,244,380) in the first degree (i.e., parent-children, siblings). One nuclear family was randomly selected from each extended family to ensure all nuclear families were independent. To balance the estimation and computational efficiency, nuclear families with at most one missing parent and up to four members were included. For families with more than two children, two children were randomly sampled for further analysis. This process yielded 129,322 patients from 40,666 nuclear families for studying the heritability of Parkinson's disease. Similar data processing applies to continuous phenotypes and integrated data for jointly analyzing two phenotypes (two-phenotype analysis). 

Inverse probability weighting (IPW) was used to address the missingness. Two important covariates were constructed to mitigate  bias due to missing: the Charlson comorbidity index (CCI)\citep{charlson} and ZIP code tabulation areas (divided into nine divisions by Census Bureau \citep{zip_code}
). The CCI predicts the mortality for a patient who may have a range of comorbidities (e.g., heart disease, AIDS, or cancer). Adjusting for the CCI can account for the differences in the health risk among patients with different comorbidities. Adjusting for the ZIP code regions may partially mitigate bias due to geographic factors, socioeconomic status, environmental exposures, healthcare access, and health behaviors. When the phenotypes include two continuous phenotypes, a multinomial model was fitted with the missingness pattern (both phenotypes missing, one phenotype missing, or none missing) as the outcome and the covariates as patients' age, gender, race, ethnicity, Charlson comorbidity index, and zip code areas. The reciprocal of the predicted probability of the observed individuals is as the individual's inverse probability weight. We randomly chose a parent and took its inverse probability weight to represent the family-specific weight. In the two-mixed-phenotype study, the phenotypes included one continuous phenotype. Hence, the IPW model simplifies into a logistic model, with the outcome being whether the continuous phenotype is observed. 

We analyzed the heritability of 6 LONIC-coded continuous laboratory tests and 12 Phecode-mapped binary phenotypes and the coheritability of 72 pairwise combinations, focusing on phenotypes related to metabolic/endocrine measurements, cardiovascular/musculoskeletal measurements, and neurological/mental disorders. Age, sex, race, and ethnicity were included as covariates in the analysis.  

\begin{figure}[htbp]
  \centering
  \begin{subfigure}[b]{0.45\textwidth}
    \centering
    \includegraphics[width=\textwidth]{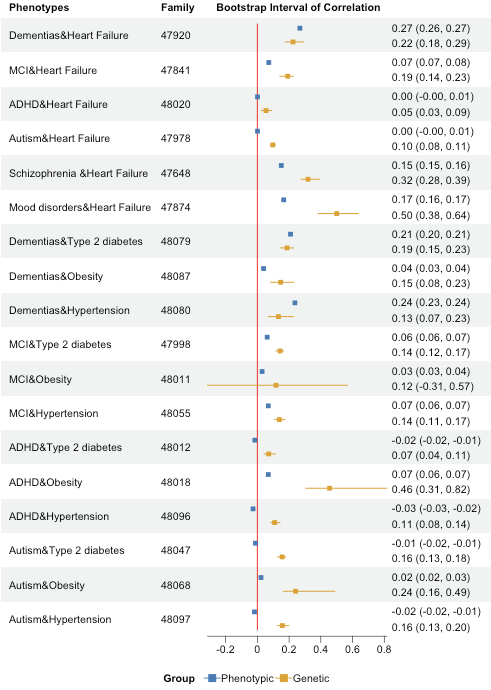}
    %\caption{Caption for left plot}
  \end{subfigure}
  %\hfill
  \begin{subfigure}[b]{0.465\textwidth}
    \centering
    \includegraphics[width=\textwidth]{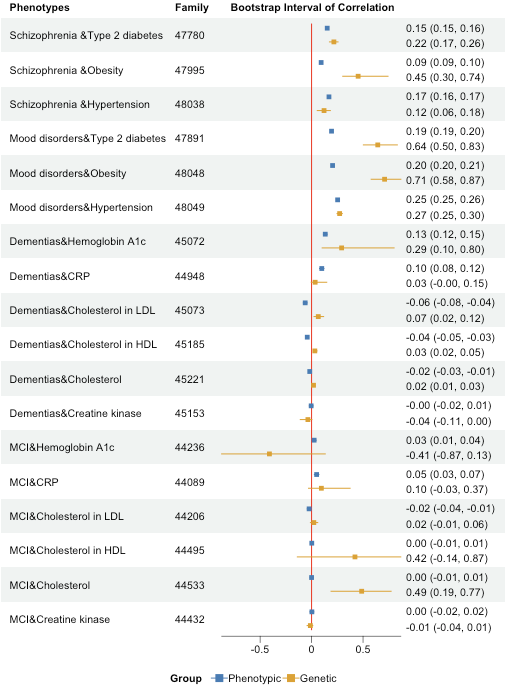}
    %\caption{Caption for right plot}
  \end{subfigure}
  \caption{Results of the point estimates and bootstrap intervals of phenotypic correlation and genetic correlation between phenotypes; percentile bootstrap is used for CI estimation; correlations were between mental and neurological diseases and metabolic, endocrine, and circulatory diseases.}
  \label{fig:forrest_plot_cor_1}
\end{figure}

Results of genetic correlation estimation are presented in Figures \ref{fig:forrest_plot_cor_1}. Additional results can be found in Figure \ref{fig:forrest_plot_cor_2} and heritability estimates are in Figure \ref{fig:forrest_plot_h}
The coheritability analysis identified several significant genetic correlations between metabolic/endocrine measures and neurological/mental disorders. For example, the genetic correlations of obesity with dementia and Parkinson’s disease were estimated at 0.147 (95\% CI: 0.083-0.228) and 0.118 (95\% CI: 0.062-0.186), respectively. These findings align with Charisis et al \citep{charisis2023obesity}, who reported that obesity may influence the expression of genes associated with Alzheimer’s disease, suggesting a genetic link between obesity and dementia.
Similarly, we observed significant genetic correlations between type 2 diabetes and both dementia (0.187, 95\% CI: 0.147-0.227) and Parkinson’s disease (0.135, 95\% CI: 0.090-0.187). These results are consistent with prior studies \citep{ji2016shared, chung2021functional, liu2023polygenic}, which suggest that type 2 diabetes and dementia, particularly Alzheimer’s disease, may share common genetic underpinnings.
Significant genetic correlations were also observed between schizophrenia and several endocrine/metabolic diseases. For instance, we found positive genetic correlations with hypertension (0.121, 95\% CI: 0.055 to 0.183), consistent with findings reported by \citep{chen2022cardiometabolic}. Positive genetic correlations were in type 2 diabetes (0.217, 95\% CI: 0.174 to 0.257), heart failure (0.320, 95\% CI: 0.275 to 0.390), and obesity (0.452, 95\% CI: 0.303 to 0.743).
These findings, however, appear to contrast with prior studies. For example, Baranova et al \citep{baranova2023shared} reported a negative genetic correlation between schizophrenia and type 2 diabetes in both European and East Asian populations. Similarly, Hartwig et al \citep{hartwig2022bidirectional} found a negative genetic correlation between schizophrenia and body mass index (BMI), suggesting an inverse genetic relationship. These discrepancies highlight the complexity of the genetic architecture underlying comorbidity and indicate the need for further investigation, potentially accounting for factors such as ancestry, phenotype definition, and analytical methods.
In addition, we observed a borderline significant genetic correlation between CRP levels and mood disorders (0.163, 95\% CI: -0.004 to 0.820), aligning with prior evidence that inflammation, metabolic dysregulation, and body mass index may play a role in the onset and severity of depression and related mood disorders\citep{kappelmann2021dissecting}.

Although metabolic phenotypes are often viewed as modifiable risk factors that can be altered through lifestyle interventions such as diet and physical activity, their potential genetic correlations with neurological and psychiatric disorders are frequently overlooked. Our findings suggest a shared genetic basis between metabolic/endocrine phenotypes and mental or neurological disorders, offering new insights into the mechanisms underlying their comorbidity. This, in turn, may inform the development of more effective, genetically informed, and personalized prevention and treatment strategies.

\section{Discussions}
Heritability and coheritability are key measures for quantifying the proportion of phenotypic variation attributable to genetic factors and the genetic correlation between distinct phenotypes. However, accurately estimating these parameters can be challenging, particularly when dealing with complex phenotypes derived from large pedigrees or massive datasets such as electronic health records (EHRs).

In this study, we propose robust statistical methods for estimating heritability and coheritability that explicitly account for familial correlations and computational constraints inherent in large-scale data. Through extensive simulation studies under various settings, we demonstrate the consistency and inferential validity of our methods. Additionally, we compare our joint modeling approach with traditional univariate analyses, showing that jointly analyzing multiple phenotypes provides improved statistical power.

Applying our methods to EHR data, we find evidence of significant genetic correlations between mental disorders and metabolic/endocrine phenotypes. These findings underscore the potential of coheritability analysis to shed light on the etiology, diagnostic strategies, and treatment development for these comorbid conditions, highlighting the need for further investigation into their shared genetic architecture.

Our approach has several limitations. When genetic correlation is close to the boundary, numerical stability issues exist in our algorithms for solving the covariance equation systems. Thus, the bootstrap confidence interval may be wider. A solution to consider is using asymptotic estimation based on estimating equations. Furthermore, our analysis focuses on the genetic correlation. Still, it highlights the need for further investigation of the causal mechanisms and pathways that link mental disorders and metabolic/endocrine measurements, as well as the potential environmental and lifestyle factors that may modulate these associations. 

Several extensions can be considered. We can extend our methods to include individuals' genotypes in the computation of SNP-based heritability when genotypes are available, or replace the familial relationships constructed from EHRs by a SNP-based genetic relatedness matrix. Furthermore, our methods can be applied to other large-scale data sources, such as biobanks and consortia, to estimate heritability and coheritability of various phenotypes and diseases and to identify potential pleiotropic effects. For example, we can use our methods to analyze the UK Biobank data, which contains genetic and phenotypic information of about 500,000 individuals from the UK. We can also consider the US-based ``All{\it of}Us" study, which includes comprehensive information on one million individuals representing the US population.

\begin{figure}
    \centering
    \includegraphics[width=.8\textwidth]{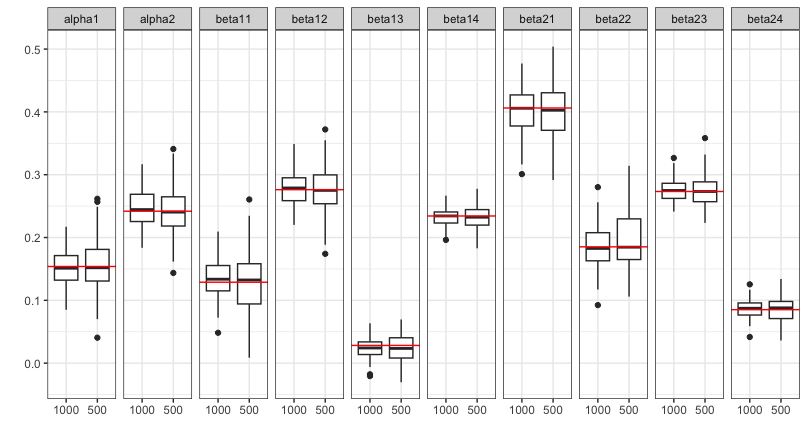}
    \includegraphics[width=.8\textwidth]{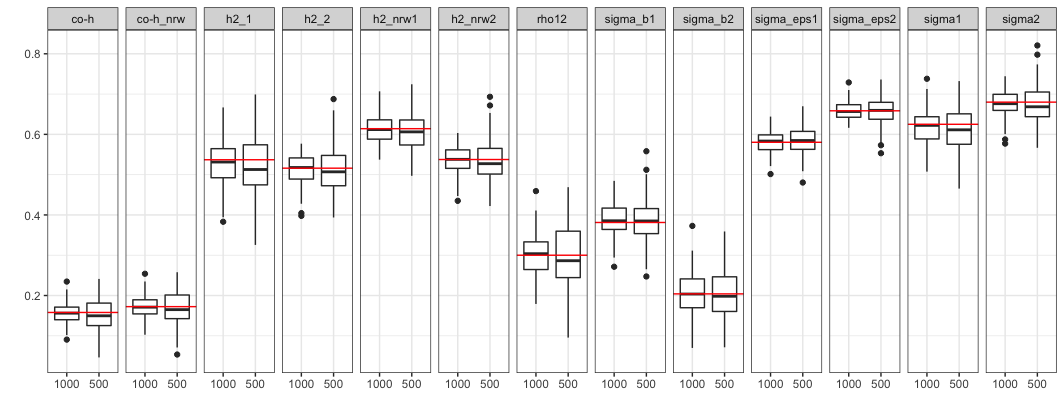}
    \caption{Parameter estimation distribution for two-continuous-phenotype analysis; results were based on 100 replicates; heritability of phenotype 1 and phenotype 2 and genetic correlation were high, high, low, respectively; simulations were conducted for 500 families and 1000 families; red lines represent true values.}
    \label{fig:continuous,high,high,low}
    \end{figure}

     \begin{figure}
    \centering
    \includegraphics[width=.8\textwidth]{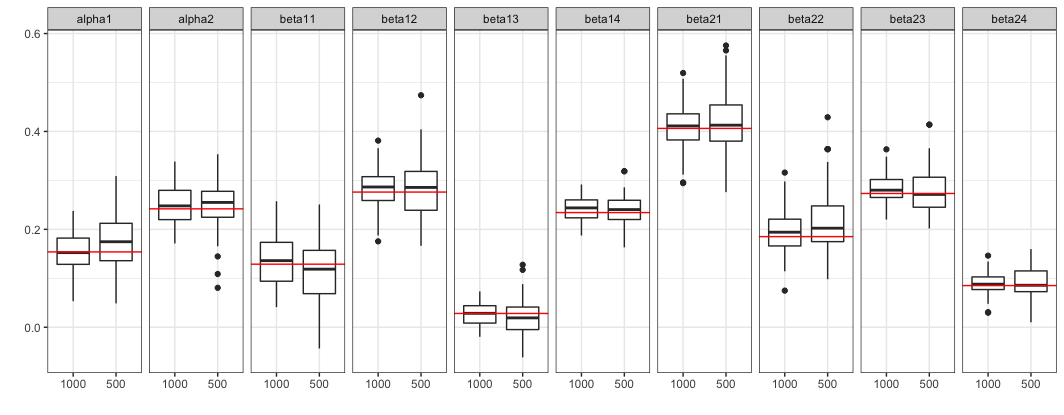}
    \includegraphics[width=.8\textwidth]{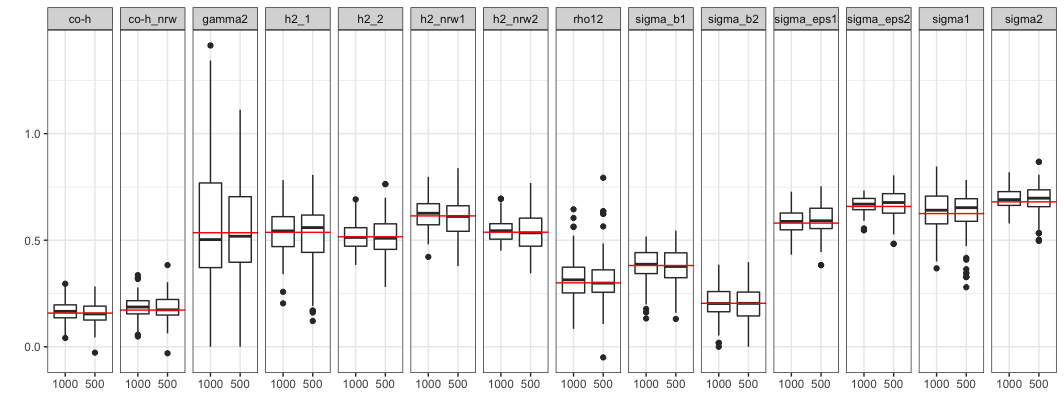}
    \caption{Parameter estimation distribution for two-binary-phenotype analysis; results were based on 100 replicates; heritability of phenotype 1 and phenotype 2 and genetic correlation were high, high, low, respectively; simulations were conducted for 500 families vs. 1000 families; red lines represent true values.}
    \label{fig:binary,high,high,low}
    \end{figure}

 \begin{figure}
    \centering
    \includegraphics[width=.8\textwidth]{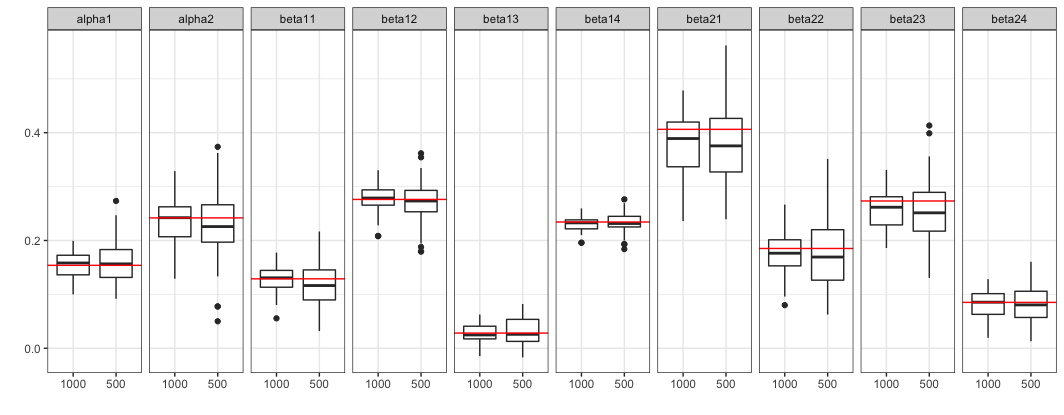}
    \includegraphics[width=0.8\textwidth]{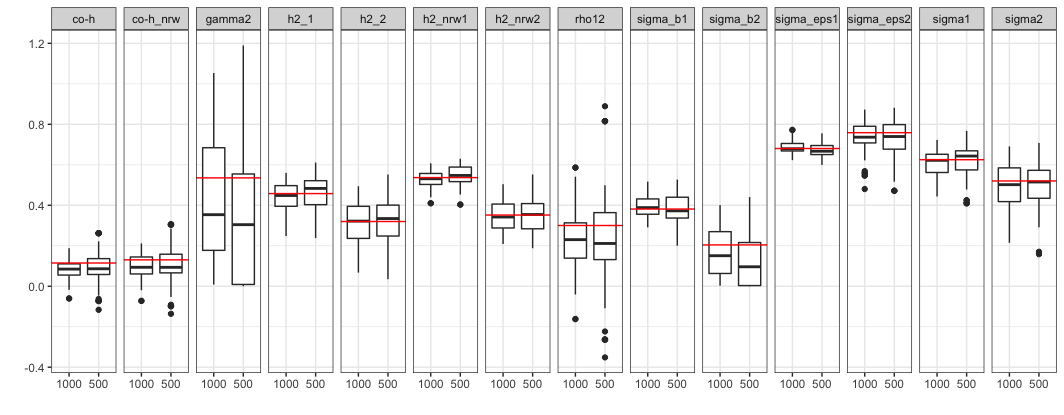}
    \caption{Parameter estimation distribution for two-mixed-phenotype analysis; results were based on 100 replicates; heritability of phenotype 1 and phenotype 2 and genetic correlation were high, high, low, respectively; simulations were conducted for 500 families vs. 1000 families; red lines represent true values.}
    \label{fig:mixed,high,high,low}
    \end{figure}

\begin{figure}%{r}{0.5\textwidth}
    \centering
\includegraphics[width=0.6\textwidth]{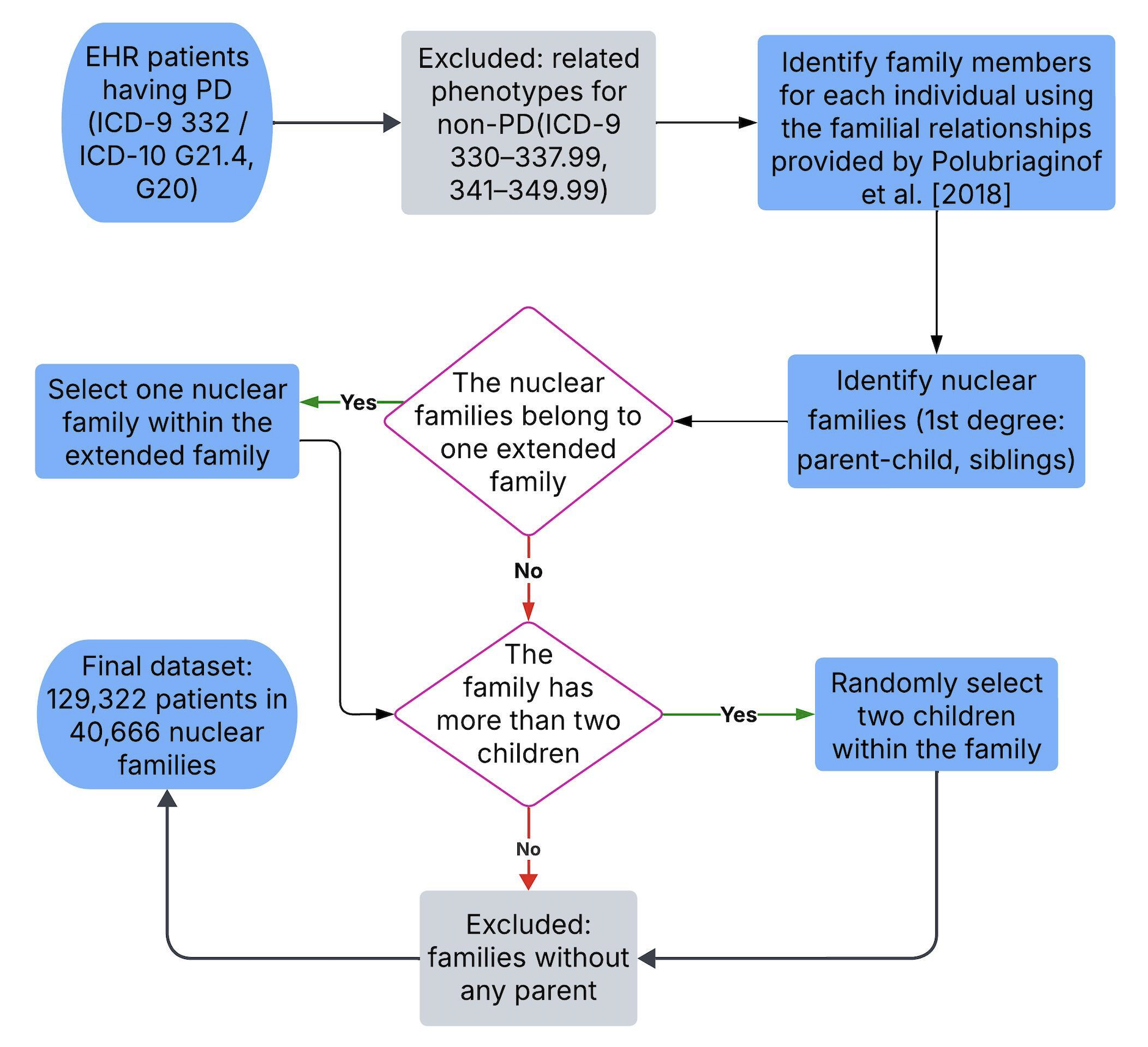}
\caption{Phenotype extraction and cohort construction illustrated with Parkinson’s Disease for the analysis of the EHR data; data is from
Columbia University Irving Medical Center in New York City EHRs.}
     \label{fig:flowchart}
\end{figure}

% \usepackage{multirow}
\begin{table}[]
\caption{Simulation results and sensitivity analysis results from two-phenotypes analysis based on 100 replicates; RMSE percent differences are calculated for 500 and 1000 families and for two-phenotype analysis(integrated) vs. single-phenotype analysis (separated); In sensitivity analysis, RMSE percent differences are calculated for families without reporting errors versus with 5\% or 10\% reporting errors.}
\label{table:RMSE}
\resizebox{\textwidth}{!}{
\begin{tabular}{clccclcccccccc}
\hline
\multicolumn{2}{c}{\textbf{Two phenotypes Type}} &
  \multicolumn{4}{c}{\textbf{Continuous}} &
  \multicolumn{4}{c}{\textbf{Binary}} &
  \multicolumn{4}{c}{\textbf{Mix}} \\ \hline
\multicolumn{2}{l}{\textbf{\begin{tabular}[c]{@{}l@{}}Heritability   of phenotype 1, \\      Heritability of phenotype 2, \\      Genetic Correlation\end{tabular}}} &
  \textbf{\begin{tabular}[c]{@{}c@{}}High,\\       High, \\      Low\end{tabular}} &
  \textbf{\begin{tabular}[c]{@{}c@{}}Low,\\       High, \\      Low\end{tabular}} &
  \textbf{\begin{tabular}[c]{@{}c@{}}Low, \\      Low, \\      Low\end{tabular}} &
  \multicolumn{1}{c}{\textbf{\begin{tabular}[c]{@{}c@{}}High, \\      High, \\      High\end{tabular}}} &
  \textbf{\begin{tabular}[c]{@{}c@{}}High,\\       High, \\      Low\end{tabular}} &
  \textbf{\begin{tabular}[c]{@{}c@{}}Low,\\       High, \\      Low\end{tabular}} &
  \textbf{\begin{tabular}[c]{@{}c@{}}Low, \\      Low, \\      Low\end{tabular}} &
  \textbf{\begin{tabular}[c]{@{}c@{}}High, \\      High, \\      High\end{tabular}} &
  \textbf{\begin{tabular}[c]{@{}c@{}}High,\\       High, \\      Low\end{tabular}} &
  \textbf{\begin{tabular}[c]{@{}c@{}}Low,\\       High, \\      Low\end{tabular}} &
  \textbf{\begin{tabular}[c]{@{}c@{}}Low, \\      Low, \\      Low\end{tabular}} &
  \textbf{\begin{tabular}[c]{@{}c@{}}High, \\      High, \\      High\end{tabular}} \\ \hline
\multirow{4}{*}{\textbf{\begin{tabular}[c]{@{}c@{}}Integrated   \\      vs.\\       Separated\end{tabular}}} &
  \textbf{Beta} &
  0.1 &
  0.3 &
  0.5 &
  0.5 &
  -1.9 &
  -1.4 &
  -2.9 &
  -0.5 &
  -0.4 &
  -1.3 &
  -0.8 &
  -2 \\ \cline{2-14} 
 &
  \textbf{Theta} &
  14.9 &
  11.3 &
  13.8 &
  11.9 &
  15.3 &
  12.5 &
  13.7 &
  11 &
  0.5 &
  4.2 &
  2.8 &
  2.3 \\ \cline{2-14} 
 &
  \textbf{Heritability} &
  9.8 &
  5 &
  9.9 &
  7.1 &
  11.3 &
  8.3 &
  12.8 &
  13.2 &
  0.2 &
  2.5 &
  2.2 &
  1.5 \\ \cline{2-14} 
 &
  \textbf{Average} &
  7.9 &
  5.7 &
  7.2 &
  6.3 &
  7 &
  3.9 &
  9.7 &
  7.8 &
  0.2 &
  1.4 &
  1 &
  1.6 \\ \hline
\multirow{4}{*}{\textbf{
\begin{tabular}[c]{@{}c@{}}1000   families\\      vs. \\      500 families\end{tabular}}} &
  \textbf{Beta} &
  28.2 &
  29 &
  28.9 &
  28.1 &
  32.7 &
  28.8 &
  28.7 &
  29.2 &
  29.1 &
  27.1 &
  31.8 &
  30.9 \\ \cline{2-14} 
 &
  \textbf{Theta} &
  26.9 &
  28.2 &
  26.5 &
  30.0 &
  34.1 &
  28.3 &
  36 &
  35.3 &
  23 &
  23 &
  27.3 &
  24.5 \\ \cline{2-14} 
 &
  \textbf{Heritability} &
  29.5 &
  31.5 &
  26.2 &
  29.9 &
  31.8 &
  27.4 &
  35.7 &
  31 &
  16.2 &
  18.1 &
  23.2 &
  20.2 \\ \cline{2-14} 
 &
  \textbf{Average} &
  28 &
  29.5 &
  27.6 &
  29.7 &
  34.2 &
  28.7 &
  33.9 &
  31.8 &
  24 &
  23.4 &
  27.9 &
  26 \\ \hline
\multirow{4}{*}{\textbf{\begin{tabular}[c]{@{}c@{}} 0 vs. 5\%   \\      Reporting Error\end{tabular}}} &
  \textbf{Beta} &
  \multicolumn{1}{l}{3.08} &
  \multicolumn{1}{l}{3.3} &
  \multicolumn{1}{l}{1.14} &
  2.25 &
  \multicolumn{1}{l}{1.67} &
  \multicolumn{1}{l}{-3.42} &
  \multicolumn{1}{l}{-0.2} &
  \multicolumn{1}{l}{11} &
  \multicolumn{1}{l}{2.25} &
  \multicolumn{1}{l}{-8.59} &
  \multicolumn{1}{l}{-1.2} &
  \multicolumn{1}{l}{8.11} \\ \cline{2-14} 
 &
  \textbf{Theta} &
  \multicolumn{1}{l}{1.33} &
  \multicolumn{1}{l}{0.06} &
  \multicolumn{1}{l}{0.99} &
  -0.2 &
  \multicolumn{1}{l}{15} &
  \multicolumn{1}{l}{0.4} &
  \multicolumn{1}{l}{-9.7} &
  \multicolumn{1}{l}{7.62} &
  \multicolumn{1}{l}{-3.94} &
  \multicolumn{1}{l}{-4.95} &
  \multicolumn{1}{l}{-3.6} &
  \multicolumn{1}{l}{-0.25} \\ \cline{2-14} 
 &
  \textbf{Heritability} &
  \multicolumn{1}{l}{-2.73} &
  \multicolumn{1}{l}{-1.48} &
  \multicolumn{1}{l}{0.42} &
  -4.19 &
  \multicolumn{1}{l}{15.2} &
  \multicolumn{1}{l}{-0.5} &
  \multicolumn{1}{l}{-15} &
  \multicolumn{1}{l}{7.09} &
  \multicolumn{1}{l}{-0.24} &
  \multicolumn{1}{l}{-5.24} &
  \multicolumn{1}{l}{2.36} &
  \multicolumn{1}{l}{-0.23} \\ \cline{2-14} 
 &
  \textbf{Average} &
  \multicolumn{1}{l}{1.36} &
  \multicolumn{1}{l}{1.57} &
  \multicolumn{1}{l}{1.21} &
  0.54 &
  \multicolumn{1}{l}{8.7} &
  \multicolumn{1}{l}{-0.99} &
  \multicolumn{1}{l}{-6.8} &
  \multicolumn{1}{l}{8.82} &
  \multicolumn{1}{l}{-0.86} &
  \multicolumn{1}{l}{-6.26} &
  \multicolumn{1}{l}{-1.6} &
  \multicolumn{1}{l}{3.57} \\ \hline
\multirow{4}{*}{\textbf{\begin{tabular}[c]{@{}c@{}}0 vs. 10\%   \\      Reporting Error\end{tabular}}} &
  \textbf{Beta} &
  \multicolumn{1}{l}{6.73} &
  \multicolumn{1}{l}{5.28} &
  \multicolumn{1}{l}{4.48} &
  4.91 &
  \multicolumn{1}{l}{1.58} &
  \multicolumn{1}{l}{0.12} &
  \multicolumn{1}{l}{-0.9} &
  \multicolumn{1}{l}{-1.69} &
  \multicolumn{1}{l}{-0.79} &
  \multicolumn{1}{l}{6.23} &
  \multicolumn{1}{l}{4.17} &
  \multicolumn{1}{l}{-0.63} \\ \cline{2-14} 
 &
  \textbf{Theta} &
  \multicolumn{1}{l}{-1.66} &
  \multicolumn{1}{l}{2.6} &
  \multicolumn{1}{l}{-1.5} &
  3.27 &
  \multicolumn{1}{l}{14.3} &
  \multicolumn{1}{l}{-5.45} &
  \multicolumn{1}{l}{-3.2} &
  \multicolumn{1}{l}{3.85} &
  \multicolumn{1}{l}{-0.02} &
  \multicolumn{1}{l}{4.86} &
  \multicolumn{1}{l}{6.96} &
  \multicolumn{1}{l}{12.4} \\ \cline{2-14} 
 &
  \textbf{Heritability} &
  \multicolumn{1}{l}{-3.26} &
  \multicolumn{1}{l}{3.58} &
  \multicolumn{1}{l}{-0.8} &
  1.66 &
  \multicolumn{1}{l}{11.1} &
  \multicolumn{1}{l}{-1.88} &
  \multicolumn{1}{l}{-1.7} &
  \multicolumn{1}{l}{7.81} &
  \multicolumn{1}{l}{-6.16} &
  \multicolumn{1}{l}{1.66} &
  \multicolumn{1}{l}{10.1} &
  \multicolumn{1}{l}{12.3} \\ \cline{2-14} 
 &
  \textbf{Average} &
  \multicolumn{1}{l}{1.66} &
  \multicolumn{1}{l}{4.32} &
  \multicolumn{1}{l}{1.6} &
  4.25 &
  \multicolumn{1}{l}{7.37} &
  \multicolumn{1}{l}{-1.94} &
  \multicolumn{1}{l}{-2.3} &
  \multicolumn{1}{l}{2.44} &
  \multicolumn{1}{l}{-2.21} &
  \multicolumn{1}{l}{4.47} &
  \multicolumn{1}{l}{6.55} &
  \multicolumn{1}{l}{6.88} \\ \hline
\end{tabular}}
\end{table}

\begin{table}[tb]
\caption{95\% CI coverage rate by parametric bootstrap; heritability of phenotype 1 and phenotype 2 and genetic correlation are high, high, low, respectively; simulations were replicated 100 times  for 1000 families; 100 bootstrap samples are included in each replicate.}
\label{table:ci rate}
\resizebox{\textwidth}{!}{
\begin{tabular}{@{}lcccccccccc@{}}
\toprule
\textbf{Two   phenotypes Type} &
  \textbf{alpha1} &
  \textbf{beta11} &
  \textbf{beta12} &
  \textbf{beta13} &
  \textbf{beta14} &
  \textbf{alpha2} &
  \textbf{beta21} &
  \textbf{beta22} &
  \textbf{beta23} &
  \textbf{beta24} \\ \midrule
Continuous & 96.0 & 92.5 & 91.5 & 93.0 & 93.5 & 94.5 & 95.5  & 93.0 & 94.5 & 95.5 \\ \midrule
Binary     & 98.1 & 95.5 & 93.6 & 97.5 & 93.6 & 94.3 & 96.2  & 94.9 & 90.5 & 94.9 \\ \midrule
Mixed      & 95.0 & 93.5 & 94.5 & 91.5 & 94.0 & 96.5 & 91.5  & 95.0 & 97.0 & 94.0 \\ \midrule
\textbf{Two phenotypes Type} &
  \textbf{gamma2} &
  \textbf{sigma1} &
  \textbf{sigma2} &
  \textbf{sigma\_b1} &
  \textbf{sigma\_eps1} &
  \textbf{sigma\_eps2} &
  \textbf{rho12} &
  \textbf{sigma\_b2} &
  \textbf{h2\_1} &
  \textbf{h2\_2} \\ \midrule
Continuous & 96.0 & 95.5 & 95.5 & 95.0 & 95.0 & 94.5 & 94.5  & 97.0 & 95.0 & 95.0 \\ \midrule
Binary     & 94.3 & 97.5 & 94.9 & 96.8 & 96.8 & 93.6 & 93.6  & 94.3 & 96.2 & 94.3 \\ \midrule
Mixed      & 99.0 & 91.5 & 98.5 & 97.0 & 92.5 & 99.0 & 100.0 & 99.0 & 93.0 & 98.5 \\ \hline
\end{tabular}}
\end{table}

%\backmatter
\bmsection*{Author contributions}

This is an author contribution text. This is an author contribution text. This is an author contribution text. This is an author contribution text. This is an author contribution text.

\bmsection*{Acknowledgments}
This work is funded in part by MH123487, NS073671 and TL1TR001875.  

\bmsection*{Financial disclosure}

None reported.

\bmsection*{Conflict of interest}

The authors declare no potential conflict of interests.

\bibliography{reference}

\bmsection*{Supporting information}

Additional supporting information may be found in the
online version of the article at the publisher’s website.

\appendix
\bmsection{Details about solving the $V(\theta) = \widehat{cov(Y)}$ equation system}

Solving the equation system $V(\theta)=\widehat{cov(Y)}$ is equivalent to solve the equation systems as below. 
\begin{eqnarray*}
Var(Y_{\cdot jk})&=&\gamma_k^2\sigma_b^2+\sigma_k^2+\sigma_{\epsilon k}^2, \\
Cov(Y_{\cdot jk}, Y_{\cdot jm})&=&\gamma_k\gamma_m\sigma_b^2+\rho_{km}\sigma_k\sigma_m, \\
Cov(Y_{\cdot jk}, Y_{\cdot sk})&=&\gamma_k^2\sigma_b^2+c_{js} \sigma_k^2, \\
Cov(Y_{\cdot jk}, Y_{\cdot ms})&=&\gamma_k\gamma_m\sigma_b^2+c_{js}\rho_{km}\sigma_k\sigma_m.
\end{eqnarray*}
We give two scenarios to explain it. 
\subsection{Scenario 1 }
\begin{figure}
  \centering
  \includegraphics[width=0.4\textwidth]{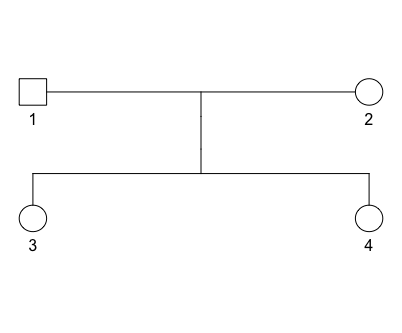}
  \caption{A two-generation family pedigree diagram showing parental and offspring relationships}
  \label{fig:pedigree}
\end{figure}

Suppose $K=1, n_i=4$. The family pedigree is shown in figure \ref{fig:pedigree}.
Let
\begin{eqnarray*}
\widehat{Cov(Y_{\cdot j1}, Y_{\cdot s1})}&=&
a_0, (j,s) \in \mathcal{A}_0,\\
\frac{1}{5} \sum_{(j,s)\in \mathcal{A}_1}\widehat{Cov(Y_{\cdot j1}, Y_{\cdot s1})}&=&
a_1,\\
\frac{1}{4}\sum_{j=1} ^4 \widehat{Var(Y_{\cdot j1})}&=&a_2\\
\end{eqnarray*}

The equation system has 3 equations with 3 parameters $(\sigma_1, \sigma_b,  \sigma_{\epsilon_1})$.

Solving this equation system can be converted into minimizing the objective function
\begin{eqnarray*}
f( \sigma_1, \sigma_b,  \sigma_{\epsilon_1}) 
= (\gamma_1^2\sigma_b^2-a_0)^2+(\gamma_1^2\sigma_b^2+1/2 \sigma_1^2-a_1)^2 
+(\gamma_1^2\sigma_b^2+\sigma_1^2+\sigma_{\epsilon 1}^2-a_2)^2.
\end{eqnarray*}

For families with full four members, the ratio of the relationship between parent and parent, parent and child, child and child, and the member itself is 1:4:1:4. The weighted objective function is
\begin{eqnarray*}
f( \sigma_1, \sigma_b,  \sigma_{\epsilon_1}) 
= (\gamma_1^2\sigma_b^2-a_0)^2+5(\gamma_1^2\sigma_b^2+1/2 \sigma_1^2-a_1)^2 
+4(\gamma_1^2\sigma_b^2+\sigma_1^2+\sigma_{\epsilon 1}^2-a_2)^2.
\end{eqnarray*}

\subsection{Scenario 2}
Suppose $K=2, n_i=4$, for each $i$. The family pedigree is the same as in scenario 1.

\begin{eqnarray*}
Var(Y_{\cdot j1})&=&\gamma_1^2\sigma_b^2+\sigma_1^2+\sigma_{\epsilon 1}^2, j=1\cdots,4.  \\
Var(Y_{\cdot j2})&=&\gamma_2^2\sigma_b^2+\sigma_2^2+\sigma_{\epsilon 2}^2, j=1\cdots,4.  \\
Cov(Y_{\cdot j1}, Y_{\cdot j2})&=&\gamma_1\gamma_2\sigma_b^2+\rho_{12}\sigma_1\sigma_2,j=1\cdots,4.  \\
Cov(Y_{\cdot j1}, Y_{\cdot s1})&=&
\begin{cases}
\gamma_1^2\sigma_b^2+1/2 \sigma_1^2, (j,s)\in \mathcal{A}_1, \\
\gamma_1^2\sigma_b^2,(j,s) \in \mathcal{A}_0
\end{cases}\\
Cov(Y_{\cdot j2}, Y_{\cdot s2})&=&
\begin{cases}
\gamma_2^2\sigma_b^2+ 1/2 \sigma_2^2, (j,s)\in \mathcal{A}_1, \\
\gamma_2^2\sigma_b^2,(j,s)\in \mathcal{A}_0
\end{cases}\\
Cov(Y_{\cdot j1}, Y_{\cdot s2})&=&
\begin{cases}
\gamma_1\gamma_2\sigma_b^2+ 1/2\rho_{12}\sigma_1\sigma_2, (j,s)\in \mathcal{A}_1,\\
\gamma_1\gamma_2\sigma_b^2, (j,s)\in \mathcal{A}_0.
\end{cases}
\end{eqnarray*}
where $\mathcal{A}_1=\{(j,s):c_{js}=1/2\}, 
\mathcal{A}_0=\{(j,s):c_{js}=0\}$. 
The equation system has 9 equations with 6 parameters $(\gamma_2, \sigma_1, \sigma_2, \sigma_b,  \sigma_{\epsilon_1}\sigma_{\epsilon_2})$. 
Let $$
a_1 =\frac{1}{4} \sum_{j=1} ^{4} \widehat{Var(Y_{\cdot j1})}, a_2 =\frac{1}{4} \sum_{j=1} ^{4} \widehat{Var(Y_{\cdot j2})},a_3 =\frac{1}{4} \sum_{j=1} ^{4} \widehat{Cov(Y_{\cdot j1}, Y_{\cdot j2})},$$
$$a_4 =\frac{1}{5}\sum_{(j,s)\in \mathcal{A}_1} \widehat{Cov(Y_{\cdot j1}, Y_{\cdot s1})},
a_5 = \widehat{Cov(Y_{\cdot j1}, Y_{\cdot s1})} ,{(j,s)\in \mathcal{A}_0},$$
$$a_6 =\frac{1}{5}\sum_{(j,s)\in \mathcal{A}_1} \widehat{Cov(Y_{\cdot j2}, Y_{\cdot s2})},
a_7 = \widehat{Cov(Y_{\cdot j2}, Y_{\cdot s2})},{(j,s)\in \mathcal{A}_0},$$
$$a_{8} =\frac{1}{5}\sum_{(j,s)\in \mathcal{A}_1} \widehat{Cov(Y_{\cdot j1}, Y_{\cdot s2})},
a_{9} = \widehat{Cov(Y_{\cdot j1}, Y_{\cdot s2})}, {(j,s)\in \mathcal{A}_0},$$

Then we have 
\begin{eqnarray*}
a_1=\gamma_1^2\sigma_b^2+\sigma_1^2+\sigma_{\epsilon 1}^2, \\ 
a_2=\gamma_2^2\sigma_b^2+\sigma_2^2+\sigma_{\epsilon 2}^2, \\  
a_3=\gamma_1\gamma_2\sigma_b^2+\rho_{12}\sigma_1\sigma_2, \\
a_4=\gamma_1^2\sigma_b^2+1/2 \sigma_1^2,  
a_5=\gamma_1^2\sigma_b^2,  \\
a_6=\gamma_2^2\sigma_b^2+ 1/2 \sigma_2^2,  
a_7=\gamma_2^2\sigma_b^2,  \\
a_8=\gamma_1\gamma_2\sigma_b^2+ 1/2\rho_{12}\sigma_1\sigma_2,
a_9=\gamma_1\gamma_2\sigma_b^2.
\end{eqnarray*}
Solving this equation system can be converted into minimizing the objective function 
\begin{eqnarray*}
&&f(\sigma_1, \sigma_2, \sigma_{b_1}, \sigma_{b_2}, \sigma_{\epsilon_1}, \sigma_{\epsilon_2}) \\
&=& 4(\sigma_{b_1}^2+\sigma_1^2+\sigma_{\epsilon 1}^2-a_1)^2
+4(\sigma_{b_2}^2+\sigma_2^2+\sigma_{\epsilon 2}^2-a_2)^2
+4(\sigma_{b_1}\sigma_{b_2}+\rho_{12}\sigma_1\sigma_2-a_3)^2\\
&+&5(\sigma_{b_1}^2+1/2 \sigma_1^2-a_4)^2 
+(\sigma_{b_1}^2-a_5)^2\\
&+&5(\sigma_{b_2}^2+ 1/2 \sigma_2^2-a_6)^2
+(\sigma_{b_2}^2-a_7)^2\\
&+&5(\sigma_{b_1}\sigma_{b_2}+ 1/2\rho_{12}\sigma_1\sigma_2-a_8)^2
+(\sigma_{b_1}\sigma_{b_2} -a_9)^2. 
\end{eqnarray*}

\bmsection{Details about estimating $cov(Y)$ in Algorithm 2}
In Algorithm 2, element of $cov(Y)$ can be written as
\begin{eqnarray*}
cov(Y_{\cdot jk}, Y_{\cdot sm}) &=& E(cov(Y_{\cdot jk}, Y_{\cdot sm}|X_{\cdot jk}, X_{\cdot sm}, Z_{\cdot jk}, Z_{\cdot sm})) \\
&+& cov(E(Y_{\cdot jk}|X_{\cdot jk}, Z_{\cdot jk},Z_{\cdot sm}), E(Y_{\cdot sm}|X_{\cdot sm}, Z_{\cdot jk}, Z_{\cdot sm}))\\
&=& E[E(Y_{\cdot jk}, Y_{\cdot sm}|X_{\cdot jk}, X_{\cdot sm}, Z_{\cdot jk}, Z_{\cdot sm})-E(Y_{\cdot jk}|X_{\cdot jk}, Z_{\cdot jk}, Z_{\cdot sm})*E(Y_{\cdot sm}|X_{\cdot sm}, Z_{\cdot jk}, Z_{\cdot sm})]\\
&+& E[(E(Y_{\cdot jk}|X_{\cdot jk}, Z_{\cdot jk}, Z_{\cdot sm})-X_{\cdot jk} \beta_{jk})(E(Y_{\cdot sm}|X_{\cdot sm}, Z_{\cdot jk}, Z_{\cdot sm})-X_{\cdot sm} \beta_{sm})]\\
&=& E[E(Y_{\cdot jk}, Y_{\cdot sm}|X_{\cdot jk}, X_{\cdot sm}, Z_{\cdot jk}, Z_{\cdot sm})-
E(Y_{\cdot jk}|X_{\cdot jk}, Z_{\cdot jk}, Z_{\cdot sm})*X_{\cdot sm} \beta_{sm}\\
&&-E(Y_{\cdot sm}|X_{\cdot sm}, Z_{\cdot jk}, Z_{\cdot sm})*X_{\cdot jk} \beta_{jk}]\\
&+& E[X_{\cdot jk} \beta_{jk} X_{\cdot sm} \beta_{sm}]\\
&=& I + II
\end{eqnarray*}

where $I$ can be estimated by the sample mean of $E(Y_{ijk}, Y_{i sm}|X_{ijk}, X_{ism}, Z_{ik}, Z_{ism})-
E(Y_{ijk}|X_{ijk}, Z_{i jk}, Z_{ism})*X_{ism} \beta_{sm}
-E(Y_{ism}|X_{ism}, Z_{i jk}, Z_{i sm})*X_{ijk} \beta_{jk}$, and $II$ can be estimated by the sample mean of $X_{ijk} \beta_{jk} X_{ism} \beta_{sm}, i=1, \cdots, N$. \\

\textbf{Remarks} 
\begin{enumerate}
    \item In stratified sampling cases, 
$I$ and $II$ are estimated by the weighted sample mean, respectively. 
\item In the case of a mix of continuous and binary outcomes,
\begin{align*}
cov(Y_{\cdot jk}, Y_{\cdot sm}) =\begin{cases}
E[(Y_{\cdot jk}-X_{\cdot jk}\beta_{jk})(Y_{\cdot sm}-X_{\cdot sm}\beta_{sm})] \quad  \text{if } Y_{\cdot\cdot k} \text{ and } Y_{\cdot\cdot m}\text{ are both continuous,}\\
I + II\quad  \text{if } Y_{\cdot\cdot k} \text{ and } Y_{\cdot\cdot m}\text{ are both binary,}\\
III\quad \text{o.w.}
\end{cases}
\end{align*}

Suppose the $k$th outcome is continuous and the $m$th outcome is binary, i.e. $Y_{\cdot jk}, Z_{\cdot sm}$ is observed, and $Y_{\cdot sm}$ is latent, then
\begin{eqnarray*}
III&=&E(cov(Y_{\cdot jk}, Y_{\cdot sm}|X_{\cdot jk}, X_{\cdot sm}, Z_{\cdot sm})) + cov(E(Y_{\cdot jk}|X_{\cdot jk}, Z_{\cdot sm}), E(Y_{\cdot sm}|X_{\cdot sm}, Z_{\cdot sm}))\\
&=&E[E(Y_{\cdot jk}, Y_{\cdot sm}|X_{\cdot jk}, X_{\cdot sm}, Z_{\cdot sm}) - E(Y_{\cdot jk}|X_{\cdot jk},  Z_{\cdot sm})*E(Y_{\cdot sm}| X_{\cdot sm}, Z_{\cdot sm})] \\
&+& E[(E(Y_{\cdot jk}|X_{\cdot jk}, Z_{\cdot sm})-X_{\cdot jk} \beta_{jk})(E(Y_{\cdot sm}|X_{\cdot sm}, Z_{\cdot sm})-X_{\cdot sm} \beta_{sm})]\\
&=& E[E(Y_{\cdot jk}, Y_{\cdot sm}|X_{\cdot jk}, X_{\cdot sm}, Z_{\cdot sm})-
E(Y_{\cdot jk}|X_{\cdot jk}, Z_{\cdot sm})*X_{\cdot sm} \beta_{sm}\\
&&-E(Y_{\cdot sm}|X_{\cdot sm}, Z_{\cdot sm})*X_{\cdot jk} \beta_{jk}]\\
&+& E[X_{\cdot jk} \beta_{jk} X_{\cdot sm} \beta_{sm}]\\
&=& (I) + (II),
\end{eqnarray*} 
where $(I)$ can be estimated by the sample mean of $E(Y_{ijk}, Y_{i sm}|X_{ijk}, X_{ism}, Z_{ism})-
E(Y_{ijk}|X_{ijk}, Z_{i sm})*X_{ism} \beta_{sm}
-E(Y_{ism}|X_{ism}, Z_{i sm})*X_{ijk} \beta_{jk}$, and $(II)$ can be estimated by the sample mean of $X_{ijk} \beta_{jk}X_{i sm} \beta_{sm}, i=1, \cdots, N$.
\item Calculate
$E[Y_{ijk}|Z_{ijk}, X_{ijk}, \beta, \theta]$, $E[Y_{ijk}|Z_{ism}, X_{ijk}, \beta, \theta]$,$E[Y_{ijk}Y_{ism}| Z_{ism}, X_{ijk},X_{ism}, \beta, \theta]$, $E[Y_{ijk}|Z_{ijk},Z_{ism}, X_{ijk}, \beta, \theta]$, and $E[Y_{ijk}Y_{ism}|Z_{ijk}, Z_{ism}, X_{ijk},X_{ism}, \beta, \theta]$
    \begin{eqnarray*}
    E[Y_{ijk}|Z_{ijk}=1, X_{ijk}=x_{ijk}]
    &=& \int_{y\in \mathbf{R}} y P(Y_{ijk}=y|Z_{ijk}=1, X_{ijk}=X_{ijk})dy\\
    &=& \int_{y\in \mathbf{R}} y\frac{P(Z_{ijk}=1|Y_{ijk}=y,X_{ijk}=x_{ijk})\cdot P(Y_{ijk}=y|X_{ijk}=x_{ijk})}{\int P(Z_{ijk}=1|Y_{ijk}=y,X_{ijk}=x_{ijk})\cdot P(Y_{ijk}=y|X_{ijk}=x_{ijk}) dy}dy
    \end{eqnarray*}
    Note that $P(Z_{ijk}=1|Y_{ijk}=y,X_{ijk}=x_{ijk})=I(y>0)$, then we have $$E[Y_{ijk}|Z_{ijk}=1, X_{ijk}=x_{ijk}]= \frac{\int_{y>0} y f_{Y_{ijk}}(y) dy}{\int_{y>0} f_{Y_{ijk}(y)}dy}$$

    Thus, \begin{eqnarray*}
    E[Y_{ijk}|Z_{ijk}, X_{ijk}, \beta, \theta]&=& 
    \begin{cases}
    \frac{\int_{y>0} y f_{Y_{ijk}}(y) \,dy}{\int_{y>0} f_{Y_{ijk}(y)}\,dy} \quad \text{if } Z_{ijk}=1,\\
    \frac{\int_{y<0} y f_{Y_{ijk}}(y) \,dy}{\int_{y<0} f_{Y_{ijk}(y)}\,dy} \quad \text{if } Z_{ijk}=0\\
    \end{cases}
    \end{eqnarray*}
        where $Y_{ijk}|X_{ijk}=x_{ijk}, \beta, \theta\sim  N(\alpha_k+\beta_k^T x_{ijk}, \gamma_k^2\sigma_b^2+\sigma_k^2+\sigma_{\epsilon k}^2)$. \\
        
    Similarly, 
    \begin{eqnarray*}
    E[Y_{ijk}^2|Z_{ijk}, X_{ijk}=x_{ijk}, \beta, \theta]&=& 
    \begin{cases}
    \frac{\int_{y>0} y^2 f_{Y_{ijk}}(y) \,dy}{\int_{y>0} f_{Y_{ijk}(y)}\,dy} \quad \text{if } Z_{ijk}=1,\\
    \frac{\int_{y<0} y^2 f_{Y_{ijk}}(y) \,dy}{\int_{y<0} f_{Y_{ijk}(y)}\,dy} \quad \text{if } Z_{ijk}=0\\
    \end{cases}
    \end{eqnarray*}
    
    \begin{eqnarray*}
E(Y_{i jk}|X_{i jk},Z_{ism}) =\begin{cases}
    \frac{\iint_{y_1 \in \mathbf{R}, y_2>0} y_1 f_{Y_{ijk},Y_{ism}}(y_1,y_2) \,dy_1\, dy_2}{\iint_{y_1 \in \mathbf{R}, y_2>0}  f_{Y_{ijk},Y_{ism}}(y_1,y_2)\,dy_1\, dy_2}\quad \text{if } Z_{ism}=1,\\
    \\
    \frac{\iint_{y_1 \in \mathbf{R}, y_2<0} y_1 y_2 f_{Y_{ijk},Y_{ism}}(y_1,y_2)\,dy_1\, dy_2}{\iint_{y_1 \in \mathbf{R}, y_2<0} f_{Y_{ijk},Y_{ism}}(y_1,y_2)\,dy_1\, dy_2}\quad \text{if } Z_{ism}=0.
    \end{cases}
\end{eqnarray*}
    
        \begin{eqnarray*}
E(Y_{i jk} Y_{i sm}|X_{i jk}, X_{ism}, Z_{i sm}) =\begin{cases}
    \frac{\iint_{y_1 \in \mathbf{R}, y_2>0} y_1 y_2 f_{Y_{ijk},Y_{ism}}(y_1,y_2) \,dy_1\, dy_2}{\iint_{y_1 \in \mathbf{R}, y_2>0} f_{Y_{ijk},Y_{ism}}(y_1,y_2)\,dy_1\, dy_2}\quad \text{if } Z_{ism}=1,\\
    \\
    \frac{\iint_{y_1 \in \mathbf{R}, y_2<0} y_1 y_2 f_{Y_{ijk},Y_{ism}}(y_1,y_2)\,dy_1\, dy_2}{\iint_{y_1 \in \mathbf{R}, y_2<0} f_{Y_{ijk},Y_{ism}}(y_1,y_2)\,dy_1\, dy_2}\quad \text{if } Z_{ism}=0.
    \end{cases}
\end{eqnarray*}
    
        \begin{eqnarray*}
    &&E[Y_{ijk}|Z_{ijk}, Z_{ism}, X_{ijk}, \beta, \theta]\nonumber\\
    &=&
    \begin{cases}
    \frac{\iint_{y_1>0, y_2>0} y_1  f_{Y_{ijk},Y_{ism}}(y_1,y_2) \,dy_1\, dy_2}{\iint_{y_1>0, y_2>0} f_{Y_{ijk},Y_{ism}}(y_1,y_2)\,dy_1\, dy_2}\quad \text{if } (Z_{ijk}, Z_{ism})=(1,1),\\
    \\
    \frac{\iint_{y_1>0, y_2<0} y_1 f_{Y_{ijk},Y_{ism}}(y_1,y_2)\,dy_1\, dy_2}{\iint_{y_1>0, y_2<0} f_{Y_{ijk},Y_{ism}}(y_1,y_2)\,dy_1\, dy_2}\quad \text{if } (Z_{ijk}, Z_{ism})=(1,0),\\
    \\
    \frac{\iint_{y_1<0, y_2>0} y_1  f_{Y_{ijk},Y_{ism}}(y_1,y_2)\,dy_1\, dy_2}{\iint_{y_1<0, y_2>0} f_{Y_{ijk},Y_{ism}}(y_1,y_2)\,dy_1\, dy_2}\quad \text{if } (Z_{ijk}, Z_{ism})=(0,1),\\
    \\
    \frac{\iint_{y_1<0, y_2<0} y_1  f_{Y_{ijk},Y_{ism}}(y_1,y_2)\,dy_1\, dy_2}{\iint_{y_1<0, y_2<0} f_{Y_{ijk},Y_{ism}}(y_1,y_2)\,dy_1\, dy_2}\quad \text{if } (Z_{ijk}, Z_{ism})=(0,0).\\
    \end{cases}
    \end{eqnarray*}

    \begin{eqnarray*}
    &&E[Y_{ijk}Y_{ism}|Z_{ijk}, Z_{ism}, X_{ijk}, X_{ism}, \beta, \theta]\nonumber\\
    &=&
    \begin{cases}
    \frac{\iint_{y_1>0, y_2>0} y_1 y_2 f_{Y_{ijk},Y_{ism}}(y_1,y_2) \,dy_1\, dy_2}{\iint_{y_1>0, y_2>0} f_{Y_{ijk},Y_{ism}}(y_1,y_2)\,dy_1\, dy_2}\quad \text{if } (Z_{ijk}, Z_{ism})=(1,1),\\
    \\
    \frac{\iint_{y_1>0, y_2<0} y_1 y_2 f_{Y_{ijk},Y_{ism}}(y_1,y_2)\,dy_1\, dy_2}{\iint_{y_1>0, y_2<0} f_{Y_{ijk},Y_{ism}}(y_1,y_2)\,dy_1\, dy_2}\quad \text{if } (Z_{ijk}, Z_{ism})=(1,0),\\
    \\
    \frac{\iint_{y_1<0, y_2>0} y_1 y_2 f_{Y_{ijk},Y_{ism}}(y_1,y_2)\,dy_1\, dy_2}{\iint_{y_1<0, y_2>0} f_{Y_{ijk},Y_{ism}}(y_1,y_2)\,dy_1\, dy_2}\quad \text{if } (Z_{ijk}, Z_{ism})=(0,1),\\
    \\
    \frac{\iint_{y_1<0, y_2<0} y_1 y_2 f_{Y_{ijk},Y_{ism}}(y_1,y_2)\,dy_1\, dy_2}{\iint_{y_1<0, y_2<0} f_{Y_{ijk},Y_{ism}}(y_1,y_2)\,dy_1\, dy_2}\quad \text{if } (Z_{ijk}, Z_{ism})=(0,0).\\
    \end{cases}
    \end{eqnarray*}
\end{enumerate}

\bmsection{Supplementary figures\&tables for real data analysis}

\begin{figure}%{r}{0.5\textwidth}
  \begin{center}
   \includegraphics[width=0.8\textwidth]{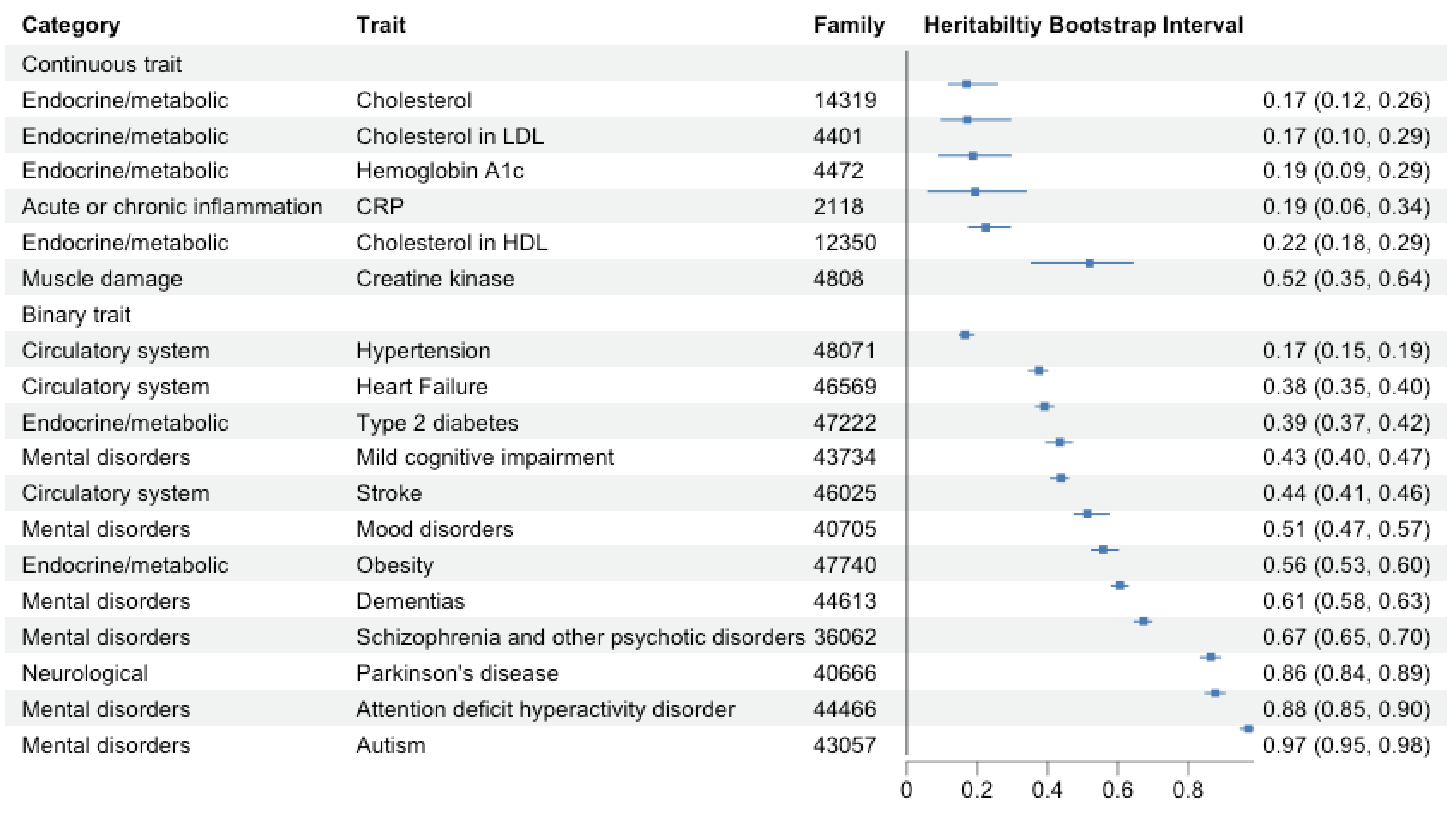}
   \caption{Point estimation and bootstrap interval of heritability $h_{12}$ for continuous biomarkers and binary phenotypes;
     ercentile bootstrap is used to obtain the  bootstrap confidence intervals.}
      \label{fig:forrest_plot_h}
  \end{center}
   
\end{figure}

\begin{figure}[htbp]
  \centering
  \begin{subfigure}[b]{0.45\textwidth}
    \centering
    \includegraphics[width=\textwidth]{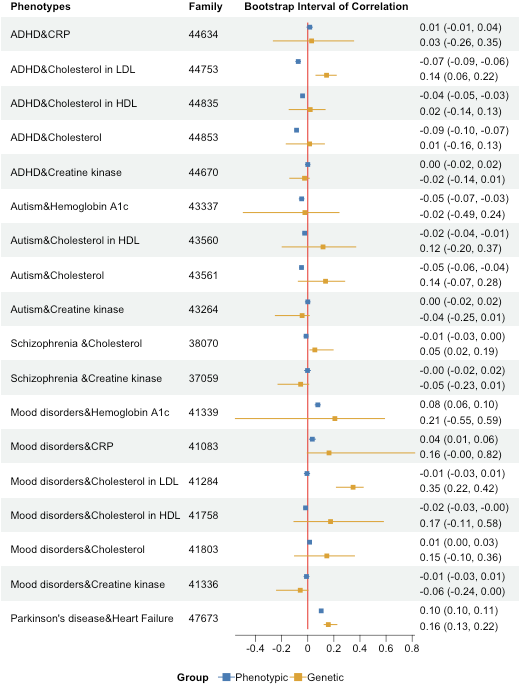}
    %\caption{Caption for left plot}
  \end{subfigure}
  %\hfill
  \begin{subfigure}[b]{0.466\textwidth}
    \centering
    \includegraphics[width=\textwidth]{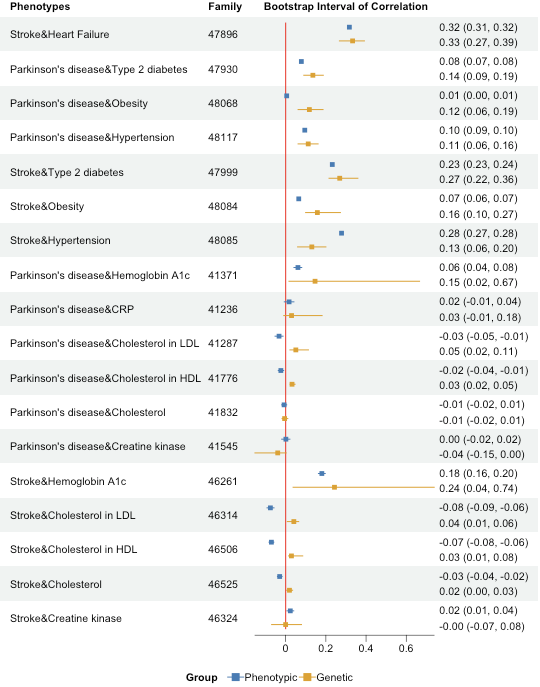}
    %\caption{Caption for right plot}
  \end{subfigure}
  \caption{Additional results of the point estimates and bootstrap intervals of phenotypic correlation and genetic correlation between phenotypes; percentile bootstrap is used for CI estimation; correlations were between mental and neurological diseases and metabolic, endocrine, and circulatory diseases.}
  \label{fig:forrest_plot_cor_2}
\end{figure}

\end{document}